\documentclass[final,3p,times]{elsarticle}

\usepackage{booktabs} 
\usepackage{tabularx} 

\usepackage{lscape}

\usepackage{amssymb}
\usepackage{pifont}
\usepackage{amsmath}
\usepackage{amsmath,amsfonts}
\usepackage{booktabs,makecell,tabularx}
\usepackage{algorithm}

\usepackage{array}
\usepackage[caption=false,font=normalsize,labelfont=sf,textfont=sf]{subfig}
\usepackage{textcomp}
\usepackage{stfloats}
\usepackage{url}
\usepackage{verbatim}
\usepackage{graphicx}
\usepackage[algo2e]{algorithm2e}
\usepackage{cite}
\usepackage{xcolor}
\usepackage{hyperref}
\usepackage{float}
\usepackage{tikz}
\usepackage{bbding}

\usepackage{pgfplots}
\usepgfplotslibrary{statistics} %
\usepackage{tabularx}
\usepackage{longtable} 
\pgfplotsset{compat=1.18}

\journal{Artificial Intelligence in Medicine}

\begin{document}

\begin{frontmatter}



\title{The Role of Machine Learning in Congenital Heart Disease Diagnosis: Datasets, Algorithms, and Insights}

 \author[1,*]{Khalil Khan}
\affiliation[1]{organization={Department of Computer Science, School of Engineering and Digital Sciences, Nazarbayev University, Kazakhstan}}
\author[2]{Ikram Syed}
\affiliation[2]{Dept of Information and Communication Engineering, Hankuk University of Foreign Studies, South Korea}

\author[3]{Farhan Ullah}
\affiliation[3]{organization={School of Internet of Things, Wuxi, University, Wuxi, China}}

\author[4]{Irfan Ullah}
\affiliation[4]{organization={School of Computer Science, Chengdu University of Technology, Chengdu, China}
}



\begin{abstract}

Congenital heart disease is among the most common fetal abnormalities and birth defects. Despite identifying numerous risk factors influencing its onset, a comprehensive understanding of its genesis and management across diverse populations remains limited. Recent advancements in machine learning have demonstrated the potential for leveraging patient data to enable early congenital heart disease detection. Over the past seven years, researchers have proposed various data-driven and algorithmic solutions to address this challenge. This paper presents a systematic review of congential heart disease recognition using machine learning, conducting a meta-analysis of 432 references from leading journals published between 2018 and 2024. A detailed investigation of 74 scholarly works highlights key factors, including databases, algorithms, applications, and solutions. Additionally, the survey outlines reported datasets used by machine learning experts for congenital heart disease recognition. Using a systematic literature review methodology, this study identifies critical challenges and opportunities in applying machine learning to congenital heart disease. 



\end{abstract}




\end{frontmatter}




\section*{Abbreviations}

\begin{flushleft}
\makebox[3cm][l]{ABNOR} Abnormal \\
\makebox[3cm][l]{$A_c$} Accuracy\\
\makebox[3cm][l]{AI} Artificial intelligence \\
\makebox[3cm][l]{AUC} Area under the curve \\
\makebox[3cm][l]{ASD} Atrial septal defect \\
\makebox[3cm][l]{BP} Back propagation \\
\makebox[3cm][l]{BCH} Beijing children's hospital \\
\makebox[3cm][l]{CDD} CirCor DigiScope dataset \\
\makebox[3cm][l]{CMR} Cardiovascular magnetic resonance \\
\makebox[3cm][l]{CHD} Congenital heart disease \\
\makebox[3cm][l]{CHD-HSY} CHD Heart Sounds Yunnan \\
\makebox[3cm][l]{CHD-ML} CHD using machine learning \\
\makebox[3cm][l]{CNN} Convolutional neural networks \\
\makebox[3cm][l]{CT} Computed tomography \\
\makebox[3cm][l]{CV} Computer vision \\
\makebox[3cm][l]{DL} Deep learning \\
\makebox[3cm][l]{$DS_c$} Dice Coefficient \\
\makebox[3cm][l]{ECG} Electrocardiography \\
\makebox[3cm][l]{Echo} Echocardiography \\
\makebox[3cm][l]{FN} False negatives \\
\makebox[3cm][l]{FCNN} fully convolutional neural network \\
\makebox[3cm][l]{$F_c$} F1-Score\\
\makebox[3cm][l]{FP} False positives \\
\makebox[3cm][l]{GAN} Generative adversarial networks \\
\makebox[3cm][l]{GB} Gradient Boosting \\
\makebox[3cm][l]{GBDT} Gradient Boosting Decision Trees \\
\makebox[3cm][l]{HSS} Heart Sounds Shenzhen Corpus \\
\makebox[3cm][l]{HSMM} Hidden Semi-Markov Model \\
\makebox[3cm][l]{HVSMR} Heart ventricles segmentation in MR \\
\makebox[3cm][l]{KNN} K-Nearest Neighbours \\
\makebox[3cm][l]{LR} Logistic regression \\
\makebox[3cm][l]{ML} Machine learning \\
\makebox[3cm][l]{MRI} Magnetic resonance imaging \\
\makebox[3cm][l]{NLP} Natural language processing \\
\makebox[3cm][l]{NN} Neural networks \\
\makebox[3cm][l]{NOR} Normal \\
\makebox[3cm][l]{PDA} Patent ductus arteriosus \\
\makebox[3cm][l]{PCG} Phonocardiogram \\
\makebox[3cm][l]{PFO} Patent foramen ovale \\
\makebox[3cm][l]{RF} Random forest \\
\makebox[3cm][l]{RLDS} Residual learning diagnosis system \\
\makebox[3cm][l]{RNN} Recurrent neural network \\
\makebox[3cm][l]{RQ} Research question \\
\makebox[3cm][l]{$R_c$} Recall\\
\makebox[3cm][l]{SLR} Systematic literature review \\
\makebox[3cm][l]{SVM} Support vector machine \\
\makebox[3cm][l]{SOA} State-of-the-art \\
\makebox[3cm][l]{$S_e$} Sensitivity\\
\makebox[3cm][l]{$S_p$} Specificity\\
\makebox[3cm][l]{TCD} The CirCor DigiScope \\
\makebox[3cm][l]{TN} True negatives \\
\makebox[3cm][l]{TNR} True negative rate \\
\makebox[3cm][l]{TOF} Tetralogy of Fallot \\
\makebox[3cm][l]{TP} True positives \\
\makebox[3cm][l]{TPR} True positive rate \\
\makebox[3cm][l]{TTE} Transthoracic echocardiography \\
\makebox[3cm][l]{TML} Traditional machine learning \\
\makebox[3cm][l]{UCI} University of California \\
\makebox[3cm][l]{US} Ultrasound \\
\makebox[3cm][l]{ViT} Vision Transformer \\
\makebox[3cm][l]{VSD} Ventricular septal defect \\

\end{flushleft}

\pagebreak
\section{Introduction}
\label{Sec-1}

CHD is a major birth defect with the highest mortality rate worldwide, with a reported incidence of approximately 1.8 per 100 live births \citep{R1,R2}. This disease can be properly cured if early detection and intervention are performed on time \citep{R3}. Moreover, a timely and accurate diagnosis is crucial for affected children to receive effective treatment, significantly increasing the likelihood of successful surgical outcomes \citep{r3, r4, r5, r6}. Medical practitioners believe that early detection and surgical intervention of CHD in a timely manner lead to an asymptomatic CHD condition, which is a healthy condition \citep{R2,Syssoyev}. However, due to a complex diagnosis process and the unavailability of pediatric cardiologists in remote areas, a significant number of critical CHDs continue to go undiagnosed \citep{r7}.

CHD encompasses cardiac abnormalities and structural alterations of the heart that are present at birth. Each person with CHD has a unique heart, made up of a unique mix of heart problems that were there before surgery and changes that happened due to long-term remodelling \citep{R0}. Researchers reported thirty-five different types of CHDs \citep{Shabana}. Nanotologists and medical practitioners have not fully explored all these types. However, most doctors can identify easily four common types: ASD, VSD, PDA, and PFO \citep{R8}. Doctors use various modalities such as echo, ECG, US, CT scan, X-rays, and POX to diagnose CHDs. To the best of our understanding, the gold standard for CHD recognition is the echo.

The adoption of AI in medical diagnostics is rapidly expanding due to its efficiency, accessibility, and operational improvements \citep{Frasca, Kumar, Kulkarni,R10, R11,Rx111,Rx112}. This growing integration has heightened the demand for AI in clinical settings. Recent developments in AI methodologies have inspired numerous studies on CHD, focusing on accelerating the scanning process, enhancing image quality, and improving diagnostic accuracy \citep{Su, Ah1, Ah}. By analyzing large datasets, AI can identify hidden patterns and relationships that might be overlooked by human experts. These insights enable AI models to deliver accurate predictions about the prevalence and severity of CHD. Researchers have also reported software applications that leverage AI for evaluating some ventricular functions through echo modality \citep{R13, R15}. Despite these advancements, we are still distant from achieving a stage where AI can entirely substitute or autonomously assist medical professionals in the identification of CHD. AI can serve as a powerful tool to complement clinicians, but achieving reliable, autonomous AI-driven diagnosis requires overcoming significant technical and ethical challenges.

\subsection{Research Questions and Key Contributions}

We conducted an SLR with 422 research publications sourced from the Scopus database. We initially analysed these publications' metadata, selecting 74 for an in-depth review. The metadata analysis aimed to address the following research questions:

\begin{itemize}

\item \textbf{RQ 1:} What are the global trends in CHD research contributions, including leading countries, institutions, and focus areas?
\item \textbf{RQ 2:} How much do funding initiatives support research integration?

\end{itemize}

Subsequently, the detailed review of 74 papers focused on the following questions:
\begin{itemize}
 \item \textbf{RQ 3:} What diagnostic methods do medical practitioners use to identify CHDs? 
 \item \textbf{RQ 4:} What are the datasets reported in the literature for CHD recognition? ion?
\item \textbf{RQ 5:} What are the most commonly used ML/DL algorithms for CHD recognition? 
\item \textbf{RQ 6:} What are the used evaluation measures, and what are the reported results on these standard datasets?
\end{itemize}

CHD-ML remains a relatively unexplored research area with limited comprehensive reviews available in the literature. Table \ref{T1} summarises some previously published review articles on CHD-ML. While most papers reviewed ML algorithms \citep{Liu,Ejaz,Pozza,Sethi,Day}, 80\% lacked a discussion on evaluation metrics—an essential component for assessing model performance. Furthermore, only 30\% of the reviewed papers included meta-analyses, highlighting a gap in systematically synthesising findings. Notably, none of the reviewed articles provided a thorough overview of CHD datasets, which would be invaluable for CV or ML experts. Additionally, many of these reviews employed traditional literature review methods, lacking adherence to SLR methodologies \citep{Marinka,Day}. Most research works also failed to specify the time-frames of the examined literature, as observed in \citep{Liu,Pozza, Jone,Hoodbhoy,Day,Marinka}. Our proposed study overcomes these limitations by adopting an SLR approach, offering a comprehensive analysis of CHD-ML encompassing datasets, algorithms, and evaluation metrics. The contributions of the proposed SLR are as follows:

\begin{table*}
\centering
\small
\caption{Survey papers reported so far on CHD-ML.}
\label{T1}
\begin{tabular}{ccccccccclc}
\toprule
\textbf{S. No.} & \textbf{Paper} & \textbf{Time-period} & \textbf{ML-Algorithms} & \textbf{Evaluation-metrics} & \textbf{meta analysis} & \textbf{Datasets} & \textbf{SLR}   \\ \midrule
1 & Liu et al. \citep{Liu} & Not specified & \ding{51} & \ding{53} & \ding{53} & \ding{53} & \ding{53}\ \\ 
2 & Ejaz et al. \citep{Ejaz} & 2015-2023 & \ding{51} & \ding{51} & \ding{53} & \ding{53} & \ding{53}  \\ 
3 & Pozza et al. \citep{Pozza}  & Not specified & \ding{51}& \ding{53} & \ding{53} & \ding{53} & \ding{53}  \\ 
5 & Sethi et al. \citep{Sethi}  &  2002-2022 & \ding{53}& \ding{53} & \ding{51} & \ding{53} & \ding{53}\\ 
6 & Jone et al. \citep{Jone}  &  Not specified & \ding{53}& \ding{53} & \ding{51} & \ding{53} & \ding{53}\\ 
7 & Hoodbhoy et al. \citep{Hoodbhoy}  &  Not specified & \ding{51}& \ding{53} & \ding{51} & \ding{53} & \ding{51}\\ 
8 & Helman et al. \citep{Helman}  &  2015-2018 & \ding{51}& \ding{53} & \ding{53} & \ding{53} & \ding{51}\\ 
9 & Day et al. \citep{Day}  &  Not specified & \ding{51}& \ding{53} & \ding{53} & \ding{53} & \ding{53}\\
10 & Marinka et al. \citep{Marinka}  &  Not specified & \ding{51}& \ding{51} & \ding{53} & \ding{53} & \ding{53}\\
\bottomrule
\end{tabular}%
\end{table*}

\begin{figure*}[t]
	\centering
     \includegraphics[width=\columnwidth,keepaspectratio]{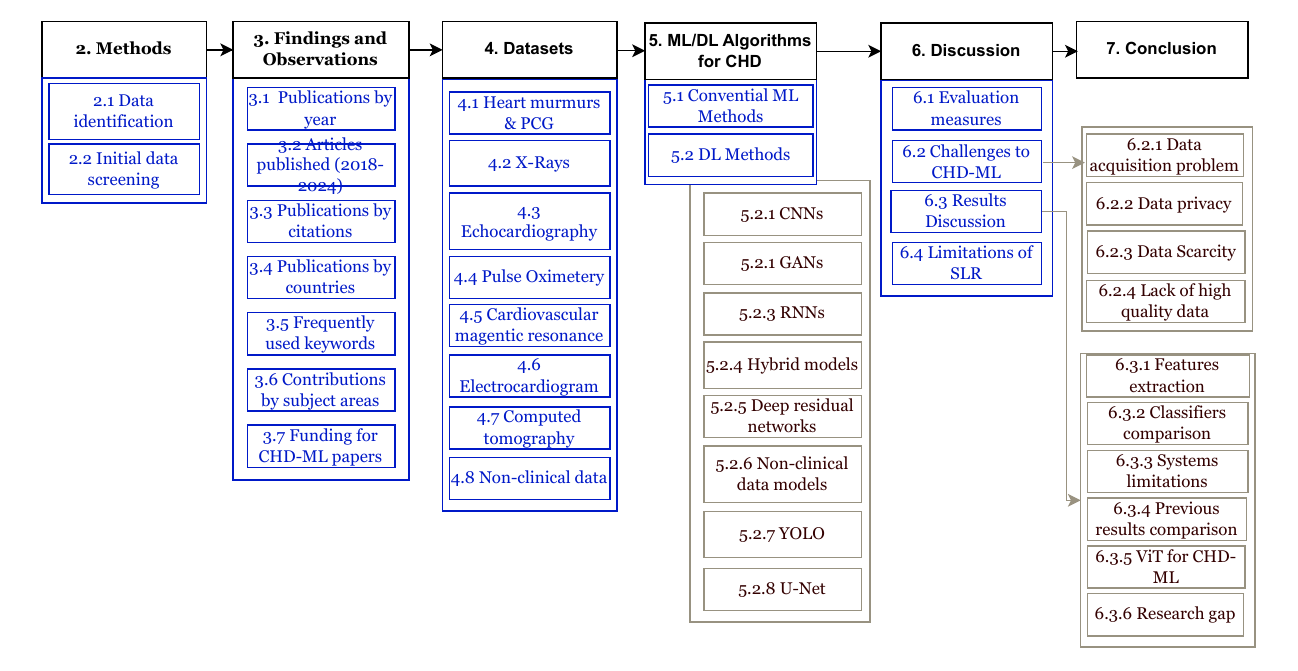}
	\caption{The structure of the proposed SLR.} 
	\vspace{0,25cm}
	\label{Fig1}
\end{figure*}

\begin{itemize}


\item Our proposed article examines all aspects of CHD-ML by referencing over 74 papers published between 2018 - 2024. We systematically review and categorize recent advancements in each domain, presenting them in a coherent and accessible manner for readers.

\item The SLR stands out as one of the most comprehensive analyses of CHD-ML, incorporating both a primary review and a targeted phase-2 review. To the best of our knowledge, it is the first SLR on CHD-ML to focus on specific research application domains and methodologies.

\item The review provides a detailed summary of the modalities used by medical practitioners and cardiovascular experts for CHD recognition. Unlike previous reviews, this paper offers a nuanced analysis of these modalities and their practical applications.

\item This study compiles and organizes all publicly available datasets relevant to CHD recognition, making it the first review to present a comprehensive inventory of datasets in this research area.

\item The paper conducts a thorough analysis of ML and DL methods applied to CHD recognition, highlighting their strengths and limitations. It spans a broad spectrum of approaches, from traditional ML techniques to the latest advancements in DL algorithms.

\item Lastly, the study identifies and discusses emerging research directions in CHD-ML, providing valuable insights for both ML/DL experts and medical practitioners interested in advancing this critical field.
\end{itemize}

We argue that CHD-ML remains an underexplored area that requires significant attention from the research community. This survey focuses on papers published within the last seven years, offering a comprehensive analysis of the key factors necessary for developing cutting-edge CHD-ML systems. The SLR provides valuable insights into the end-to-end CHD-ML framework, enabling researchers to gain a holistic understanding of the field and facilitating structured exploration and development.

\subsection{Organization of SLR}

The primary objective of this SLR is to serve as a comprehensive reference for researchers and practitioners. It aims to provide an in-depth analysis of current trends and methods while identifying research gaps to facilitate the development of advanced ML and DL-based approaches for CHD recognition. The structure of this paper is illustrated in Fig. \ref{Fig1} and organized as follows: Section \ref{sec-2} outlines the SLR methodology, observations and key findings are included in Section \ref{sec-3}. Section \ref{sec-4} details the various diagnostic methods and databases employed for CHD recognition. Section \ref{sec-5} examines ML and DL algorithms, providing insights into their applications for CHD recognition. Section \ref{sec-6} presents a comprehensive discussion and future directions. The paper conclusion is presented in Section \ref{sec-7}. This structure ensures a logical flow, guiding readers through the key components of the SLR and its findings.

\section{Methods}
\label{sec-2}

An SLR involves the formulation of questions and the application of systematic and clear methodologies to identify, select, and critically evaluate relevant research papers from the studies reported so far \citep{r022}. We select this method due to its accuracy and reliability in synthesizing academic material, as well as its widespread acceptance across several research disciplines. We perform the SLR in accordance with the PRISMA criteria. Although PRISMA is not a quality evaluation approach, it is extensively recognised in the research community for its evidence-based checklist items and four-phase analysis \citep{r023}.

\subsection{Data Identification}

\begin{figure*}[t]
	\centering
	\includegraphics[width=17cm, height=14cm]{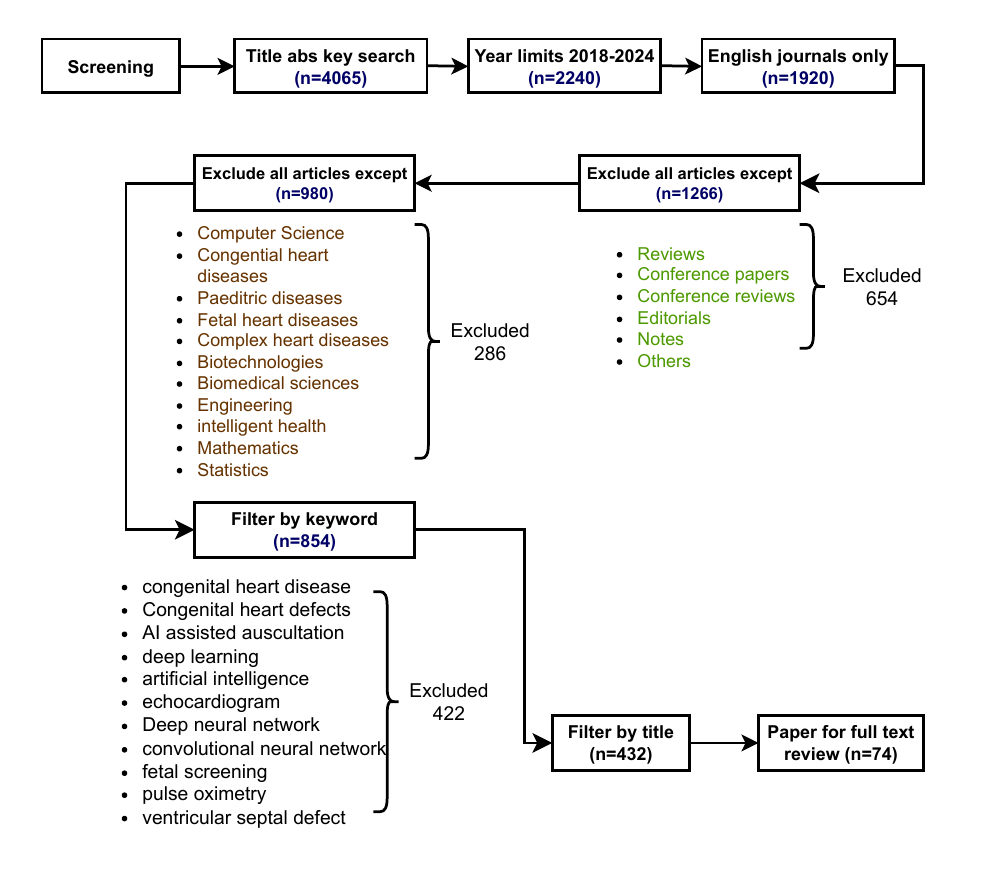}
    \caption{PRISMA flow diagram used in this research.} 
	\vspace{0,25cm}
	\label{Fig2}
\end{figure*}

We conducted thorough research utilizing the Scopus integrated database, which encompasses all principal publishers, including Emerald, Elsevier, Springer, Taylor \& Francis, IEEE, MDPI, Nature, and Wiley. The Scopus database is regarded as a reputable resource by numerous scholars for conducting SLR, owing to its high-quality indexed content \citep{r024,r025}. The search encompasses the period from January 2018 to November 2024 and incorporates all pertinent articles released throughout this timeframe. In our search for pertinent articles, we employed keywords such as “congential heart,” “machine learning,” “artificial intelligence,” and “congenital heart disease recognition.” Boolean operators are employed in conjunction with various keywords to expand the search area. The search procedure was devised. The survey includes each paper that combines empirical research with experimental results. We meticulously adhere to controlled vocabulary and associated keywords to effectively refine the search parameters. The comprehensive search approach is illustrated in Fig. \ref{Fig2}.

\subsection{Initial Data Screening}

The preliminary search in the Scopus database, utilizing specific keywords, yielded 4065 articles. Following the implementation of the year limit from 2018 to 2024, the article count dropped to 2240. The constraints of document type, language, subject area, and keywords reduce the total number of articles to 432. We identified 432 publications for screening based on their titles and abstracts after applying keywords and the search methodology. We imported data from 432 articles in Excel CSV format for further analysis. Duplicates were detected and eliminated utilizing Excel's duplicate functionality. We further reviewed the remaining 432 unique article titles and abstracts to ensure their inclusion. We evaluated 432 article titles and abstracts utilizing a standardized extraction form. We rejected studies that were unrelated to ML but pertinent to heart disease, or vice versa. After screening the titles, we thoroughly reviewed 74 articles, all of which satisfied the inclusion criteria. Figure \ref{Fig2} delineates the inclusion and exclusion criteria employed in this investigation.

\section{Findings and Observations}
\label{sec-3}

The results of the metadata analysis are presented in the next subsection. The results were obtained from a detailed analysis of 74 papers and then a metadata assessment of 432 journal articles. The metadata study included 432 papers classified by year, journals, authors, subject areas, funding sources, and institutions.

\subsection{Publication by Year}

\begin{figure*}[t]
    \centering
    \includegraphics[width=12 cm, height=6cm]{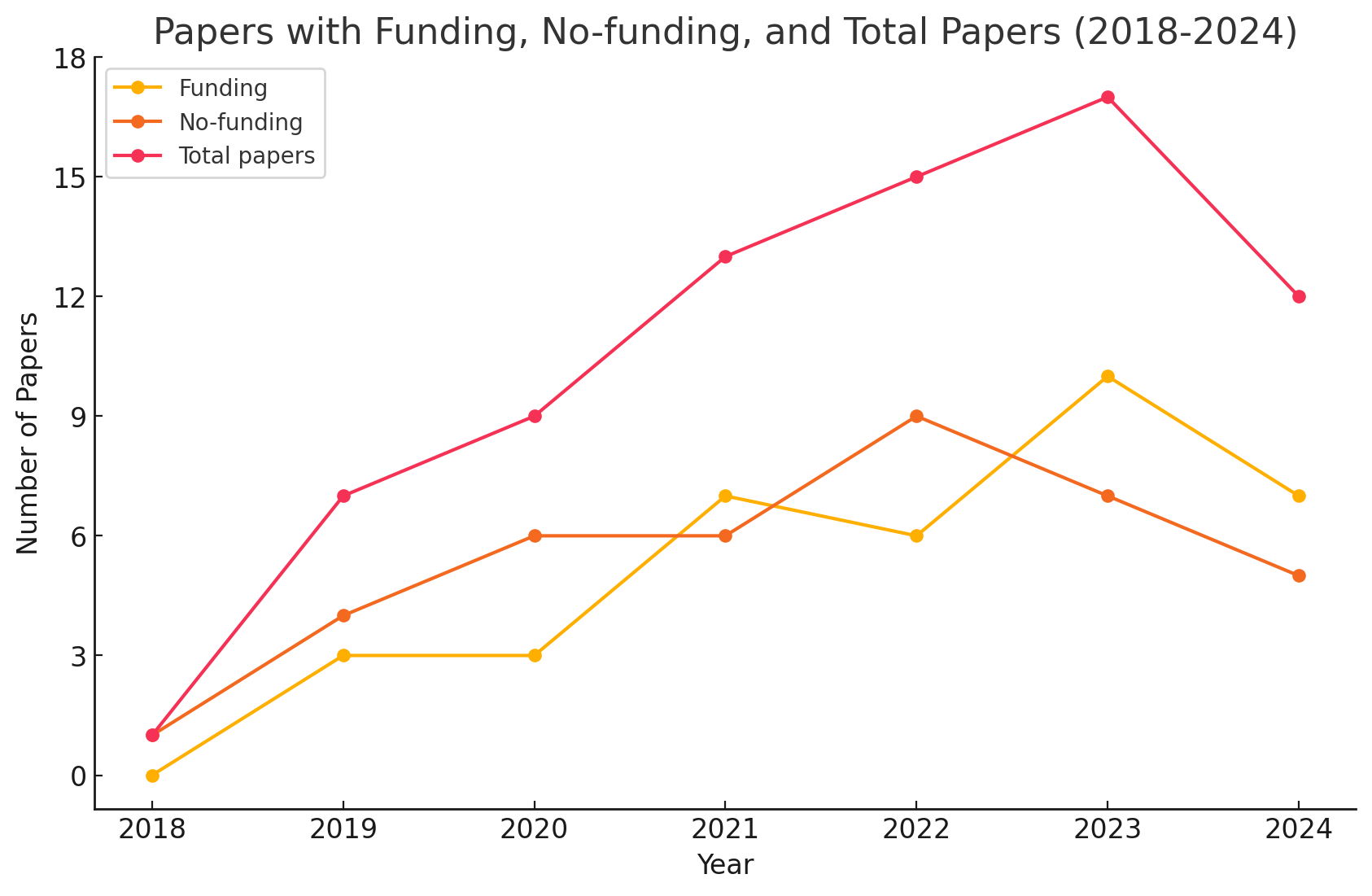}
    \caption{Total research papers published per year (2018-2024), papers with funding, and without funding.} 
    \vspace{0,25cm}
    \label{Fig3}
\end{figure*}

The annual analysis of research articles on CHD recognition from 2018 to 2024 indicates a consistent upward trend, underscoring an increasing interest in the domain. Please see Figure \ref{Fig3} for details. The total number of papers exhibits a steady rise from 2018 (1 paper) to 2023 (17 papers), thereafter seeing a minor decrease in 2024 (12 papers). We can expect some more papers in 2024 since the year is not finished and there might be papers which are under review. Publications supported by funding shown a general increase over time, peaking at 10 publications in 2023. No-funding papers experienced initial growth from 2018 to 2022 but have since stabilized at approximately 6-9 papers annually in recent years. The rising involvement of funding indicates enhanced institutional and organizational backing for CHD research utilizing ML/DL, along with the wider acknowledgment of its significance. 

\subsection{Scholarly articles published between 2018 and 2024}

CHD-ML is a critical area in healthcare where ML demonstrates promising potential. However, the volume of research papers specifically addressing CHD-ML remains relatively low compared to other healthcare applications. This scarcity of literature has resulted in minimal repetition of CHD-related ML studies within the same journals, reflecting the niche and evolving nature of this research domain. Notably, most of the high-quality papers in this field are published in journals dedicated to the medical or cardiovascular fields, highlighting the focused interest of these disciplines in advancing CHD-ML research. The rarity of repeated publications on CHD-ML in the same journal also underscores the uniqueness and diversity of these studies, as well as the broad interest across multiple outlets. We must add, the lack of repeated contributions by the same authors in CHD-related research indicates a significant gap in continuity within this field. This suggests that most researchers do not pursue long-term investigations or build on their previous work, highlighting the need for sustained focus and collaboration to advance CHD studies comprehensively.

Several notable papers on CHD-ML have been published in prestigious journals, underscoring the growing importance of this interdisciplinary field, please see Table \ref{journals} for details. For example, publications have appeared in Nature Medicine (Impact Factor: 58.7), the European Journal of Heart Failure (Impact Factor: 38.1), the European Society of Cardiology Journal (Impact Factor: 35.4), and the Journal of the American College of Cardiology (Impact Factor: 21.1). Many of these journals are among the highest-ranked in the medical and cardiovascular domains, known for their rigorous peer-review processes and high impact factors. The presence of CHD-ML research in such esteemed venues highlights its potential and emphasizes the critical need for advanced ML solutions in addressing challenges within this healthcare domain.

\begin{table*}
\centering
\caption{Top 10 journals with work on CHD-ML.}
\label{journals}
\begin{tabular}{ccccccccclc}
\toprule
\textbf{S. No.} & \textbf{Journal} & \textbf{Impact factor} \\ \midrule
1 & Nature medicine
  & 58.7 \\ 
2 & European Heart Journal
r & 38.1  \\ 
3 & European Society of Cardiology
  & 35.4 \\ 
4 & Journal of the American College of Cardiology
  & 21.7 \\ 
5 & The Lancet Child \& Adolescent Health
 & 19.9  \\ 
6 & Nature Communication
  & 14.7 \\ 
7 & Medical Image Analysis
  & 10.6 \\ 
8 & Scientific data
  & 9.8 \\ 
9 & IEEE Transactions of Medical Imaging
  & 8.9 \\ 
10 & Future Generation Computer Systems
  & 8.6 \\ 
\bottomrule
\end{tabular}%
\end{table*}

\begin{figure*}[t]
	\centering
	\includegraphics[width=15 cm, height=10cm]{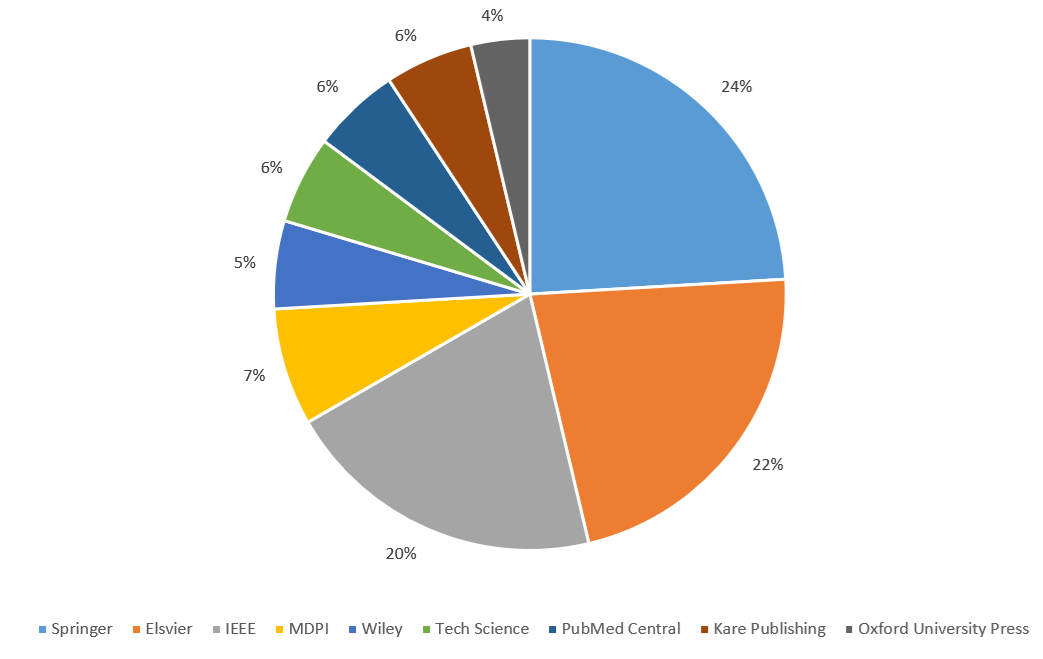}
	\caption{Publishers that contributed research publications to CHD-ML.} 
	\vspace{0,25cm}
	\label{Publishers}
\end{figure*}

\begin{table}
\centering
\small
\caption{Top 10 cited research papers published between 2018-2024 on CHD-ML.}
\label{citations}
\begin{tabular}{m{0.5cm}m{2.5cm}m{10cm}m{2cm}}
\toprule
\textbf{S. No.} & \textbf{Author} & \textbf{Title} & \textbf{total citations}\\ \midrule
1. & Liang \citep{Liang} & Evaluation and accurate diagnoses of pediatric diseases using artificial intelligence & 620\\
2. & Acharya et al. \citep{Acharya} & Deep convolutional neural network for the automated diagnosis of congestive heart failure using ECG signals
  & 326 \\ 
3. & Arnaout et al. \citep{Arnaout} & An ensemble of neural networks provides expert-level prenatal detection of complex congenital heart disease & 206\\  
4. & Oliveira et al. \citep{Oliveira} & The CirCor DigiScope dataset: from murmur detection to murmur classification &   151\\ 
5. & Pace et al. \citep{Pace} & Interactive whole-heart segmentation in congenital heart disease & 147 \\
6. & Thompson et al. \citep{Thompson} & Artificial intelligence-assisted auscultation of heart murmurs: validation by virtual clinical trial & 85\\
7. & Gong et al. \citep{Gong} & Fetal congenital heart disease echocardiogram screening based on DGACNN: adversarial one-class classification combined with video transfer learning & 71\\
8. & Liu et al. \citep{Liu2} & Classification of heart diseases based on ECG signals using long short-term memory & 69 \\
9. & Xu et al. \citep{Xu} & Whole heart and great vessel segmentation in congenital heart disease using deep neural networks and graph matching & 62\\
10. & Karimi et al. \citep{Karimi} & Fully‑automated deep‑learning segmentation of pediatric cardiovascular magnetic resonance of patients with complex congenital heart diseases & 54\\ 

\bottomrule
\centering
\end{tabular}
\end{table}

We would also like to comment here that the presence of CHD-ML studies in high-impact journals such as Nature Medicine etc. underscores the inherently multidisciplinary nature of this research. Advancing this field requires collaboration between experts in medicine, computer science, and engineering to bridge the gap between clinical needs and technological solutions. Encouraging such interdisciplinary research is essential for fostering innovation and driving the growth of impactful studies in CHD diagnosis and management using ML.

\begin{figure*}[t]
	\centering
	\includegraphics[width=\textwidth, height=7cm]{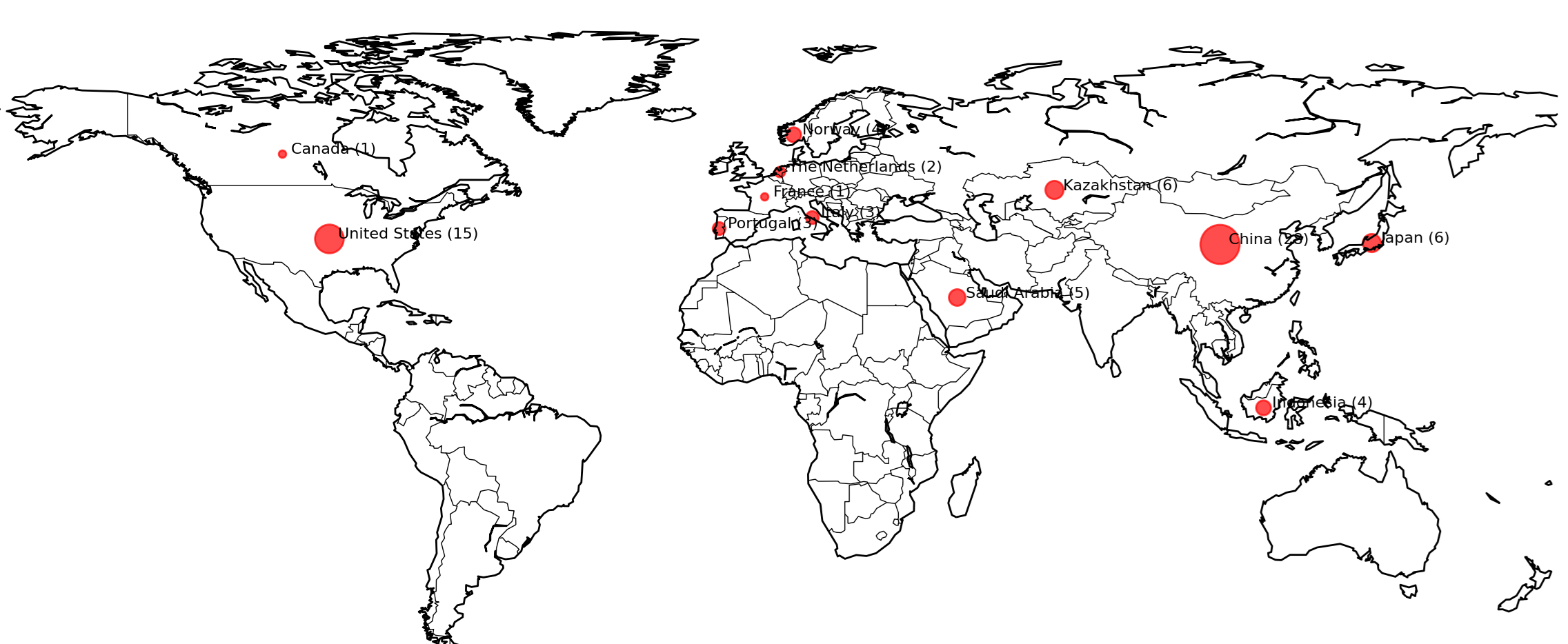}
	\caption{Countries that contributed in the form of publications to CHD-ML.} 
	\vspace{0,25cm}
	\label{Countries}
\end{figure*}

The analysis of publishers contributing to research on CHD-ML highlights a diverse distribution, as illustrated in the accompanying Figure \ref{Publishers}. Leading the contributions are Springer and Elsevier, accounting for 24\% and 22\% of the publications, respectively. This dominance indicates the central role of established academic publishers in disseminating high-quality research in CHD-ML. IEEE follows closely with 20\%, reflecting the significant contribution of engineering and technology-focused domains to this interdisciplinary field. Other notable contributors include MDPI (7\%), Wiley (6\%), and Tech Science (6\%), which collectively represent smaller but growing contributions to the CHD-ML literature.

Additionally, publishers such as PubMed Central, Kare Publishing, and Oxford University Press, though accounting for smaller shares (5\% or less), play an important role in reaching specialized audiences. The diverse representation across both medical and technical publishers emphasizes the interdisciplinary nature of CHD-ML research, requiring contributions from experts in machine learning, engineering, and medicine.

\subsection{Publications By Citations}

We do not observe frequent citations of CHD-ML in the literature. Frequent citation of research works provides a useful indicator of prominent contributors and impactful studies. Table \ref{citations} displays the ten most frequently referenced articles in the Scopus database as of November 2024. It is worth noting that citation counts may vary slightly from Google Scholar due to differences in indexing methodologies and timelines. According to the data, the publication by Liang et al., focusing on the accurate diagnosis of CHD using AI, garnered the highest citation count with 620 citations. This is followed by Acharya et al., whose work on deep CNNs for diagnosing CHD using ECG signals received 326 citations. Other notable contributions include Arnaout et al. (206 citations) and Oliveira et al. (151 citations).

An analysis of these citation trends reveals two important insights. First, the number of citations across these papers is relatively low compared to more established fields, reflecting the nascent stage of CHD-ML as a research domain. Despite the emerging nature of this field, the variety of topics and methodologies in these studies underscores its interdisciplinary potential. Second, the data highlights a pressing need for further exploration, innovation, and dissemination of impactful findings in this domain. Fostering interdisciplinary collaborations between medical and technical researchers will be crucial in driving growth, improving adoption, and positioning CHD-ML as a mature and well-recognized research area.

\subsection{Publications by Countries}

Figure \ref{Countries} illustrates the contributions of of research articles from various countries in CHD-ML. China accounts for the highest number of publications, contributing 46.3\% of the total referenced works. The United States follows as the second-largest contributor, with 24.1\%, highlighting its significant role in advancing this domain. Japan and Kazakhstan are tied for the third position, each contributing 5.6\% of the total output. Other notable contributors include Saudi Arabia, Indonesia, Norway, Italy, and Portugal, each accounting for 3.7\%, while smaller contributions come from the Netherlands, France, and Canada, each with 1.9\%. These figures demonstrate China's dominance in CHD-ML research and reflect the USA's strong but secondary role in terms of the total publication share. The contributions from Japan, Kazakhstan, and other countries underscore the global interest in advancing this interdisciplinary field. The significant participation of multiple nations suggests the growing importance of fostering international collaboration to further accelerate advancements in CHD-ML research.

\subsection{Frequently Used Keywords}

Table \ref{keywords} displays the most prevalent single, double, and triple keywords found in the papers. The most commonly utilised keywords are identified utilising the R software tool. Our initial purpose was to discover and analyse the article that exclusively focused on terms such as ML, DL. We came to know that that among single keywords, only ``Echocardiogram" and ``congential" ranked among the most often utilised terms by the authors in the titles. The term ``deep learning" is the most frequently utilised double keyword (28 occurrences), followed by ``heart failure" (22 occurrences). A triple keyword ``congential heart disease" (40 occurrences) as a triple keyword and heart sound asculation with 37 occurencies.

\begin{figure*}[t]
	\centering
	\includegraphics[width=\textwidth, height=5cm]{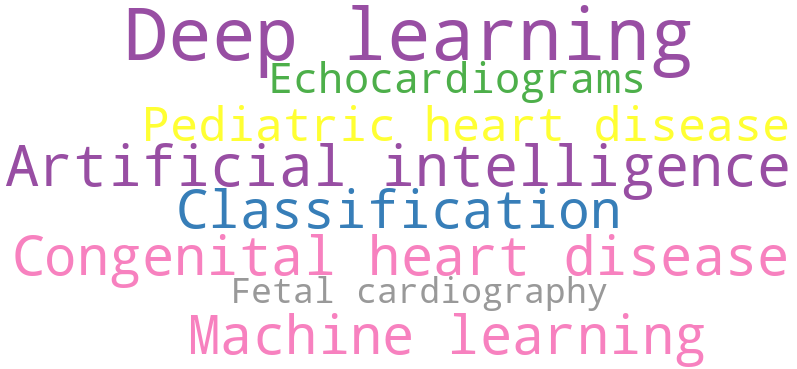}
	\caption{The word count of the most frequent key words in the titles.} 
	\vspace{0,25cm}
	\label{word-count}
\end{figure*}

\begin{table*}
\centering
\caption{Frequently used single, double, and triple keywords.}
\label{keywords}
\begin{tabular}{ccccccccclc}
\toprule
\textbf{Unigrams} & \textbf{Frequency} & \textbf{Bigrams} & \textbf{Frequency} & \textbf{Trigrams} & \textbf{Frequency} \\ \midrule
echocardiogram & 26 & deep learning
& 28 & congenital heart disease & 40 \\
Congenital & 22 & heart failure & 21 & heart sound auscultation & 37\\
heart & 16 & artificial intelligence
& 17 & Congenital heart defects
& 12\\
classification & 11 & heart failure & 14 & AI assisted auscultation & 11\\
healthcare & 10 & computer vision & 12 & Deep neural
network & 14
\\
Phonocardiography & 7 &fetal screening & 12 & cognetive heart failure & 12\\
Epidemiology & 7 & pulse oximetry & 11 & convolutional neural network & 11
\\
prediction & 8 & Computed tomography & 10 & heart sound auscultation & 10 \\
diagnosis & 9 & prenatal diagnosis & 9 & ventricular septal defect
& 8
\\
mortality & 6 & cardiovascular health & 7 & support vector machine &
7
\\
\bottomrule
\end{tabular}%
\end{table*}

An analysis of the authors' keywords in the titles of the publications revealed some interesting results. The most frequently used keywords were ``congenital heart disease,” ``machine learning,” ``deep learning,” and ``artificial intelligence,” appearing 35, 27, 17, and 12 times, respectively. These findings highlight the central themes and focus areas within CHD-ML research, emphasizing the prominence of AI-related methodologies and their application to CHD diagnosis and management. The word count of the most frequenty keywords is shown in Figure \ref{word-count}. 

\subsection{Contributions By Subject Areas}

\begin{figure}[H]
	\centering
	\includegraphics[width=15cm, height=8cm]{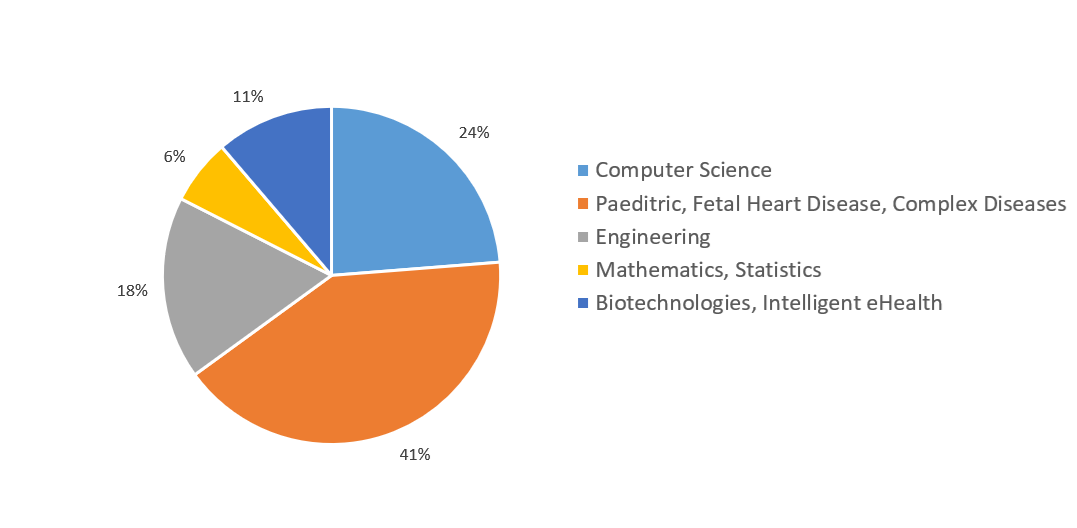}
	\caption{Publications on CHD-ML based on subject areas.} 
	\vspace{0,25cm}
	\label{F-6}
\end{figure}

Figure \ref{F-6} illustrates the distribution of research contributions across various subject areas related to CHD-ML. The most prominent domain is medicine, specifically focusing on pediatrics, fetal heart disease, and complex diseases, accounting for 41\% of the literature. Computer science follows with 24\%, reflecting the significant role of computational methods such as ML and AI in advancing CHD diagnosis and management. Engineering contributes 18\%, highlighting its relevance in developing innovative diagnostic tools and frameworks. Other fields such as mathematics and statistics (11\%) and biotechnologies, including intelligent eHealth systems (6\%), also play a crucial role in CHD-ML research. These contributions underscore the interdisciplinary nature of the field, where advancements rely on the collaboration of medicine, computer science, and engineering, along with supportive contributions from other specialized disciplines.

\subsection{Funding for CHD-ML Based On Data From Papers}

Figure \ref{Fig3} illustrates the number of research articles associated with some fundings. The quantity of publications produced from 2018 to 2024 varied about financing sources. Nevertheless, in 2022 and 2023, exponential expansion is evident, signifying that the CHD-ML attracts the interest of academics, practitioners, and funders, ultimately manifested in the rise of sponsored and published research.

\section{Datasets}
\label{sec-4} 

The effectiveness of any ML-based approach heavily depends on the dataset used for experimental validation and evaluation. This section discusses the datasets developed to date for CHD recognition. Different diagnostic modalities  are used by the medical practitioners to diagnose CHDs. The preliminary assessment for CHD in babies involves a combination of cardiac auscultation, POX, chest radiography, echo, and ECG. Diagnosing methods like X-rays, MRI of the heart, dual-source tomography scans, and CT examinations are very common; however, they are a lengthy complex process, expensive, and require qualified cardiologists \citep{r7}. It is a fact that there is a common tendency of missing diagnosis relatively late, even when the condition warrants prompt management. In the following subsections, we discuss various modalities used for CHD recognition and their corresponding datasets reported in the literature.

\subsection{Heart Murmurs and PCG Datasets}
A heart murmur is an audible sound or vibration caused by turbulent blood flow within the heart. This murmurs can occur due to a variety of reasons, such as valve abnormalities, CHDs, or increased blood flow conditions (e.g., during pregnancy or fever). Murmurs are classified as either: Innocent ( no underlying heart problem) and Abnormal (associated with structural heart issues). PCG is a tool used to record and analyze the sounds of the heart, including murmurs. A PCG device captures the heart's acoustic signals (such as normal heart sounds, S1 and S2 and ABNOR sounds such as gallops) and visualizes them in a graphical waveform. In summary, a heart murmur is a phenomenon or symptom, while PCG is a technique used to detect and analyze heart sounds, including murmurs.

AI has the potential to enhance the accuracy of auscultatory observations in the diagnosis of CHD. Heart disease recognition with ML in case of adults is somehow mature research area. A review article regarding this can be explored in \citep{Ismail}. Some research papers which use murmurs or PCG for CHD recognition can be explored in \citep{Thompson,Xu1,Liu1, Yang, Lv1, Belinha1, Ou1,Ref12}. We must add here that the auscultatory findings of CHD are essential for clinical diagnosis and offer a cost-effective tool; however, they rely on clinical expertise, which poses a limitation in resource-constrained countries, necessitating objective and reportable support for clinicians, including peripheral health workers. The deficiency of skilled cardiologists at the peripheral level results in an inevitable delay in the clinical diagnosis of CHD, leading to postponed intervention and consequently a poor prognosis \citep{RX97,RX68}. Advancements in AI for the detection of heart murmurs demonstrate potential sensitivity; nevertheless, clinical validation is necessary prior to broad clinical endorsement \citep{RX98}. Summary of the datasets for heart murmurs and PCG data is presented in Table \ref{datasets}.

\subsubsection{ZCHSound} 

The ZCHSound \citep{Ref26} was collected at the Children's Hospital of Zhejiang University (ZU) and its affiliated facilities across China. ZU Kids Hospital is a prestigious institution dedicated to providing comprehensive pediatric healthcare services. Participation in the study was entirely voluntary, with explicit consent obtained from the participants' parents. Cardiac auscultation data were recorded using an advanced stethoscope with a frequency range of 8000 Hz.

The duration of heart sound recordings for each subject ranged from 11 to 30 seconds. Acquiring heart sounds from newborns and adolescents presents distinct problems, as factors such as crying, coughing, gastrointestinal activity, and physical exercise may generate extraneous noise. To facilitate the selection of appropriate data for analysis, annotations are supplied to indicate data quality. The heart sound dataset comprises 941 people, each possessing a single audio recording. Each audio recording has an approximate length of 20 seconds, culminating in a total duration of 5 hours.

\subsubsection{CirCor DigiScope} 

The TCD \citep{Ref23} comprises heart sound recordings collected in 2014 (2014-2015). The total recordings are 5,282 obtained from 1,568 individuals, with an almost equal representation of males (787) and females (781). The participants ranged in age from infancy to young adulthood. These recordings, captured using a Littmann stethoscope, span 30 seconds each and underwent a quality assessment focused on murmur detection. For annotation accuracy, the data were reviewed by two cardiac specialist. The segmentation of heart sound samples was performed using three distinct automated approaches, with the segmentation outcomes and their implications thoroughly evaluated by the physiologists.

\subsubsection{Sounds Shenzhen Corpus}

The HSS \citep{Ref24} includes heart sound recordings obtained from 170 babies. Recordings were collected from four common sites, with each site recorded for 30 seconds. Echo analysis was performed by examining the area ratios of the mitral and tricuspid valves to assess regurgitation. Regurgitation severity was categorized into three levels: minimal, moderate, and severe. The heart sound recordings were further grouped into three categories based on severity.

\subsubsection{Heart Sounds Yunnan}

The CHD-HSY database is an extensive compilation of heart sound recordings designed to enhance the research and diagnosis of CHD. This database has about 1,892 heart sound recordings obtained from persons aged between 1 and 72 years. The data was collected from  2017 to 2019 in Yunnan Province, China. Recordings were acquired utilising electronic stethoscopes, concentrating on distinct heart sound categories linked to CHD, such as ASD, VSD, and PDA. The CHD-HSY database is a crucial resource for the development and assessment of ML models for diagnosing CHD, offering a varied collection of heart sound data for research and clinical use.

\subsubsection{DigiScope} 

The data for DigiScope \citep{Ref25} weas collected from a cohort of 29 healthy youths ( ages: 6 months-17 years). The dataset has babies heart sounds that were recorded in the Portuguese Hospital. The frequency of 4000 HZ has been kept while sounds were captured. A physician personally determined the onset and conclusion of S1 and S2 utilizing sophisticated software to examine cardiac sound.

\subsubsection{PhysioNet}

The PhysioNet \citep{LiuPCG} is widely recognized for its extensive volume and diverse age range. It comprises data from nine separate sources, including 2,435 heart sound recordings. The data is collected from 1,297 individuals where the age of individuals is  between 18 and 86 years. To standardize the data, recordings from various devices and sampling frequencies were re-sampled to 2000 Hz using appropriate filters. Heart sounds were captured from four primary auscultation sites.

\subsection{X-Rays}

Despite advancements in imaging modalities, radiography still remains one of the most commonly utilized diagnostic imaging tests for patients with CHDs. X-rays are frequently employed to detect structural abnormalities in the hearts of infants. While X-rays provide valuable initial information, they are not sufficient for a full diagnosis.

Chest radiography is widely available, easy to perform, and cost-effective, making it a practical choice for preliminary assessments. It is particularly useful in evaluating heart size, dimensions, and pulmonary blood flow, facilitating the initial diagnosis and recognition of CHD patients. This enables the timely initiation of treatment and further clinical decision-making.

However, it is important to note that X-rays has limitations in providing comprehensive information about CHDs in infants. While it serves as a useful screening tool for identifying potential abnormalities, detailed examinations and advanced imaging techniques are required for definitive diagnosis and management. Once abnormalities are detected through chest radiography, patients can be referred for more in-depth investigations using modalities such as echo or cardiac MRI.

\subsubsection{DICOM} 

The DICOM \citep{Zhixin} consists of 828 radiograph files sourced from children, categorised into four groups: ASD, VSD, PDA, and NORM. Each chest radiograph associated with a particular cardiac issue is supported with a cardiac US report. Cardiologists gathered data on paediatric patients with CHDs who experienced hospitalisations. The interval between the cardiac US report and the chest radiograph does not exceed more than 3 days. Three certified cardiologists assessed the US findings and validated the diagnosis of CHD following the first storage of chest radiographs in DICOM format.

\subsubsection{Chest X-rays}

The Chest X-rays dataset proposed \citep{Han} consists of 3,255 frontal preoperative chest radiographs gathered retrospectively from patients under 18 years old between January 2018 and February 2022. The dataset comprised 1,174 radiographs of CHD cases and 2,081 non-CHD controls, with 142 CHD cases further classified as having pulmonary arterial hypertension associated with CHD. Photographs underwent a meticulous three-tier inspection procedure to guarantee quality, eliminating illegible, duplicate, or misidentified photographs. The preprocessing processes encompassed cropping, resizing to $448\times448$ pixels, random rotation, and normalization, with data augmentation implemented to mitigate class imbalance, especially for PAH-CHD cases. 

\subsection{Echocardiography}

Echo and US are invaluable and most frequently used tools in diagnosing of CHD. Both rely on sound waves that further create images of heart internal structures, but echo is specifically tailored for the heart, offering detailed visualization of its chambers, valves, and blood flow. Through techniques like Doppler echo, clinicians can assess the direction and velocity of blood flow, helping to identify abnormalities such as septal defects, valve malformations, or improper circulation often associated with CHD. In pediatric and neonatal care, echo is particularly crucial as it provides a non-invasive, real-time assessment of structural and functional cardiac anomalies. US, more broadly, is used during prenatal screening to detect fetal CHD by evaluating the heart's development and function within the womb. Together, these imaging modalities enable early diagnosis and intervention, improving outcomes for patients with CHDs. 

To the best of our knowledge, echo is the most effective and reliable modality for the diagnosis of CHD \citep{r8, r9, r10}. Research consistently identifies echo as the gold standard for CHD diagnosis due to its ability to provide detailed and accurate cardiac imaging. However, the use of echo is labor-intensive and requires highly specialized expertise, which limits its accessibility and availability. One significant limitation is the availability of echo machines, particularly in rural or remote areas far from major cities. Additionally, expert knowledge is essential for interpreting echo results, but trained specialists are often scarce in resource-constrained settings. These challenges contribute to delays in diagnosis and limit the widespread use of this diagnostic technique. 

\subsubsection{Doppler TTE} 
Two-dimensional US Doppler techniques provide unique echo data essential to comprehending cardiac structure. CHD typically shows distinctive characteristics in both a two-dimensional and Doppler transthoracic echo, aiding its diagnosis. However, cardiac sonographers may find confirmation of CHD problematic, leading to misunderstanding and missed diagnoses. This dataset \citep{Cheng} consisted of Doppler TTE scans of about 1,932 children, obtained at BCH (2018-2022). The initial dataset included 1,080 children, comprising  healthy, ASD, andVSD. The second data set included 624 children, consisting of healthy babies, ASDs, and VSDs.  All samples for this study were given by patients at the cardiac centre. The TTE data from a cohort of 1,932 people devoid of structural cardiac abnormalities was employed as healthy control data subsequent to the examination. The status of all participants was evaluated by a minimum of two highly qualified senior US technicians and a chief physician. 

\subsubsection{Multiview CHD-echo}

Multiview CHD-echo \citep{R20} comprises a total of 1,308 children. Each patient has echo video recordings varying from 1 to 5 views. Furthermore, the physician chooses one high-quality 2D frame from each viewpoint in a film to compile a 2D echo dataset. During data acquisition, the patient was positioned supine, and the chest was exposed for the echo.  The transducer frequency varied between 3 and 8 MHz. Utilising the heart segmental technique, the locations of the heart, atria, and ventricles were determined, and the interrelations among the atria, ventricle, and aorta were examined. The atrial and ventricular septa were evaluated for anomalies, and the pulmonary venous return was analysed. Five standard two-dimensional perspectives were obtained by the authors. All patients' diagnosis were corroborated by at least two prominent ultrasonography specialists. 

\subsubsection{CHDEcho-Net}

CHDNets \citep{ref6} consists of 5840 echo recordings including three prevalent CHDs. Each echo film is associated with an individual patient and a specific visit. Keyframes in each echo film were meticulously selected by proficient specialist. Each echo was assessed using a three-tier evaluation approach. A first-level assessment was performed by a medical student possessing a bachelor's degree or above . Second-level evaluation was conducted by two junior echocardiographers, while third-level evaluation was carried out by two seasoned echocardiographers with over 10 years of clinical expertise. This tri-level evaluation approach guaranteed that each echo was accurately labelled with the appropriate diagnosis and cardiac defect localisation. Upon finalising the data labelling, 100 objects from the obtained dataset were chosen and reviewed by a third-party echocardiographist possessing over 20 years of clinical expertise. This strategy mitigate the influence and chances of human error the modelling process.

\subsubsection{Seven-CHDs}
The dataset presented in \citep{Nurmaini} consists of seven different kinds of CHDs. The research consists of fetal echo data obtained from 76 pregnant women at a hospital in Indonesia. Among these, 31 instances had CHD conditions, whereas 45 served as normal controls. The dataset comprises 1,129 frames in an intra-patient context and 55 echo for evaluation in an inter-patient context. Frames were extracted from apical four-chamber view ultrasound videos, resized from $640\times480$ pixels to $300\times300$ pixels, and annotated with seven CHD classifications. Data augmentation was conducted, expanding the dataset to 23,504 frames to improve model performance, especially in inter-patient contexts.

\subsection{Pulse Oximetery}

POX is a technique that measures the oxygen saturation level in the blood. CHD screening with POX is being progressively implemented by medical practitioners in clinics since its inception \citep{Ref38}. This strategy is also economical according to widely recognised standards \citep{Ref36,Ref37}. This approach is not especially successful on an individual basis. It is used when combined with an additional diagnostic procedure. compared to other complex methods for diagnosis, it is extremely easy to use and yields results in a surprisingly quick time frame.  POX screening can be performed easily within 24 hours for each neonate. Typically, POX screening can overlook certain types of CHDs, and significant changes in oxygen saturation can go undetected in some instances. In contrast to echo, POX offers the most efficient strategy, taking only 2-3 minutes to verify the findings \citep{Ref27}. single article using POX and ML for CHD recognition has been written by Huang et al. \citep{Ref8}.

We believe that this is not a legitimate diagnostic method for identifying CHD; nevertheless, it can be significant in identifying situations that require further investigation. This method is highly successful for detecting significant CHDs in neonates, although diminished oxygen saturation levels may require additional diagnostic assessments such as echo. However, its limitations, including the potential for FPs and FNs, underscore the need to combine POX with supplementary clinical assessments for precise CHD diagnosis. The only dataset available for CHD recognition using POX is delineated below.

\subsubsection{Pulse Oximeter Dataset} 

The POX dataset  \citep{Ref8} was gathered throughout a two-year research period. The hospital documented a total of 44,147 births. The majority of infants were delivered at full term. A total of 498 newborns with CHD were initially found; 27 were identified using pulse oximetry screening, while 471 were detected with cardiac auscultation. This led to a total screening rate of 1.13\% among the 44,147 live births. Out of the total cases, 458 neonates were assessed using ECG. This comprises 253 male newborns and 245 female infants. The principal types of CHDs included PDA (34.3\%),  ASD (20.5\%), VSD (8.3\%) and mixed anomalies of 34.5\%.





\begin{table*}[!htbp]
\centering
\captionsetup{justification=centering, singlelinecheck=off} 
\caption{Datasets reported in litrature for CHD.}
\label{datasets}
\resizebox{\textwidth}{!}{%
\begin{tabular}{@{}p{0.8cm}p{2.5cm}p{1.5cm}p{2cm}p{2cm}p{2.5cm}p{3cm}@{}}
\toprule
\textbf{S. No.} & \textbf{Dataset} & \textbf{Year} & \textbf{Applicants} & \textbf{Publicly Available?} & \textbf{Age Range} & \textbf{CHD Types} \\ 
\midrule

\multicolumn{7}{l}{\textbf{PCG and Murmurs}} \\ \midrule
1 & ZCHSound \citep{Ref26} & 2020--2022 & 1259 & \checkmark & 1 day to 14 years & NOR, ASD, VSD, PDA, PFA \\ 
2 & TCDD \citep{Ref23} & 2014--2015 & 1568 & \checkmark & 3 days to 30 years & --- \\ 
3 & HSS \citep{Ref24} & 2019 & 170 & $\times$ & 21 to 88 years & NOR, mild, severe \\ 
4 & Digiscope \citep{Ref25} & 2017 & 29 & $\times$ & 6 months to 17 years & --- \\ 
5 & PhysioNet \citep{Ref7} & 2016 & 1297 & \checkmark & 18 to 86 years & NOR, ABNOR \\ 

\multicolumn{7}{l}{\textbf{X-Rays}} \\ \midrule
6 & DICOM \citep{Ref34} & 2021--2022 & 828 & \checkmark & Not mentioned & NOR, VSD, ASD, PDA \\ 
7 & Chest X-rays \citep{Han} & 2018--2022 & 1174 & $\times$ & Under 18 years & NOR, ABNOR \\ 

\multicolumn{7}{l}{\textbf{Echo}} \\ \midrule
8 & Doppler TTE \citep{Cheng} & 2018--2022 & 1932 & $\times$ & Not available & NOR, ASD, VSD \\ 
9 & CHDEcho Net \citep{ref6} & 2023 & Not available & \checkmark & Not available & NOR, ASD, VSD \\ 
10 & Multiview CHD-Echo \citep{R20} & 2021 & 1308 & $\times$ & Not available & NOR, ASD, VSD \\ 
11 & Seven-CHDs \citep{Nurmaini} & 2022 & 76 & \checkmark & Not available & Seven CHDs \\ 

\multicolumn{7}{l}{\textbf{MRI}} \\ \midrule
12 & HVSMR2-0 & 2024 & 12 & \checkmark & Not available & NOR, ASD, VSD, PDA \\ 
13 & Complex-CHD \citep{Karimi} & 2020 & 64 & $\times$ & 1--18 years & Complex CHDs \\ 
14 & Real-Synthetic-CHDs \citep{Diller} & 2020 & 303 & $\times$ & Not available & NOR, ASD, VSD, TOF \\ 

\multicolumn{7}{l}{\textbf{ECG}} \\ \midrule
15 & CHDdECG \citep{Chen} & 2014--2023 & 65,869 & \checkmark & 2.12 ± 1.50 years & NOR, ASD, VSD, PDA, TOF \\ 
16 & PhysioNet \citep{Ref7} & 2024 & 33 & \checkmark & Not available & NOR, ABNOR \\ 
17 & Foetuses-CHD \citep{Vullings} & 2019 & 386 & $\times$ & Not available & NOR, ASD, VSD \\ 

\multicolumn{7}{l}{\textbf{POX}} \\ \midrule
18 & POX \citep{Ref36} & 2022 & 44,147 & $\times$ & Not available & NOR, ABNOR \\ 

\multicolumn{7}{l}{\textbf{Non-Clinical Datasets}} \\ \midrule
19 & Shanxi \citep{Luo2} & 2006--2008, 2020 & 33,831 & \checkmark & Not available & NOR, ABNOR \\ 
20 & UCI \citep{Ref43} & 2022 & 1050 & \checkmark & Not available & ASD, VSD, NOR \\ 
21 & Guangzhou Hospital Dataset \citep{Ref43} & 2016--2017, 2019 & 567,498 & $\times$ & 0--18 years & ABNOR, NOR \\ 

\multicolumn{7}{l}{\textbf{CT Images}} \\ \midrule
22 & ImageCHD \citep{Xu2} & --- & 110 & \checkmark & 1 month to 40 years & 16 variants of CHD \\ 

\bottomrule
\end{tabular}%
}
\end{table*}

\subsection{Cardiovascular Magnetic Resonance Datasets}

CMR is a sophisticated imaging modality designed for evaluating the anatomy (heart), functionality (heart), and vascular system. It employs magnetic fields and radiofrequency pulses to generate intricate, high-resolution 3D images. CMR imaging is essential in the identification of CHD as it visualises intricate anatomical abnormalities, evaluates cardiac functionality, and analyses haemodynamic flow patterns. It is very effective for identifying septal abnormalities, vascular anomalies, and assessing healed or palliated congenital heart diseases. CMR facilitates pre-surgical planning and post-operative surveillance, providing unparalleled clarity in the evaluation of cardiac and vascular architecture. Nonetheless, its elevated expense, restricted accessibility, and contraindications for certain patients may provide difficulties. CMR is essential for the thorough diagnosis and management of CHD, serving as a complimentary modality to echo and CT. The datasets reported for CHD recognition using CMR are reported in Table \ref{datasets}. Details of these datasets are in the following paragraphs.

\subsubsection{HVSMR-2.0}
The literature has reported HVSMR-2.0 as the latest dataset for CHD recognition. Researchers from BCH and MIT collaborated to compile the HVSMR-2.0 \citep{Pace2}. The initial version of the same database was made available to the research community in 2016 \citep{251}. The HVSMR-2.0 comprises 3D CMR images obtained during standard clinical procedures. This dataset serves as a significant resource for investigating CHD, including instances with various cardiac anomalies, particularly those that have received treatments.

The imaging utilised a 1.5T MRI scanner, applying a steady-state axial orientation. Respiratory navigation and ECG algorithms were utilised during image acquisition to reduce motion artefacts resulting from cardiac and respiratory activities. The dataset comprises hand segmentations of the blood pool and ventricular myocardium, diligently executed and verified by professional evaluators and clinical specialists.

The HVSMR-2.0 signifies a notable progression in CHD research, as it meets the demand for comprehensive cardiovascular datasets with corresponding hand segmentation masks. This dataset comprises 60 CMR scans. The images include diverse cardiac abnormalities. The dataset contains masks that define the critical and supplementary extents of primary vessels, hence improving the accuracy of comparative analyses amongst algorithms. HVSMR-2.0 provides a basis to develop clinically pertinent devices that may significantly impact the management and treatment of patients with CHDs.

\subsubsection{Complex-CHD}

The dataset reported in \citep{Karimi} shows a set of 64 CMR scans from the Children's Hospital Los Angeles of kids who had complicated CHD. The dataset encompasses a variety of heart diseases, such as TOF, double outlet right ventricle, transposition of the great arteries, cardiomyopathy, coronary artery anomalies, pulmonary stenosis/atresia, truncus arteriosus, and aortic arch anomalies. The dataset includes kids between two and 18 years old. The dataset has 20 to 30 frames per cardiac cycle and 12 to 15 short-axis slices for each subject. These slices cover both the left and right ventricles. Authors used 1.5T and 3T scanners and followed a standard procedure. No contrast agents were used and an experienced paediatric cardiologist adhered to SCMR standards to establish the ground truth, manually marking the end-diastolic and end-systolic volumes.

A deep GAN, which created synthetic CMR images and segmentation masks, expanded the dataset to address the lack of pediatric CHD data. This addition greatly expanded the database, encompassing a wider range of heart shapes and conditions. This dataset, along with new methods for augmentation and segmentation, is a big step towards making automated analysis and clinical workflows better for kids with congenital heart disease.

\subsubsection{Real-Synethetic-CHDs }

The dataset proposed in \citep{Diller} consists of 6400 long-axis MRI frames. All the patients had CHDs, particularly those with post-repair ToF, a common condition that causes the heart to turn blue. Fourteen centers in Germany participated in a state study that included MRIs. Patients over 8 years old who did not have implantable cardioverter-defibrillators met the inclusion criteria. The MRIs were saved in DICOM format and had information like the weight, height, and age. Data were made anonymous and were allowed to be used in a study in an ethical way.

Progressive GANs were also used to make synthetic MRI images. The images made by Progressive GANs were $256\times256$ pixels and showed both short- and long-axis views. These synthetic images were used to build a U-Net segmentation model to separate the left ventricle, the right ventricle, and the right atrium. This dataset demonstrates the use of GAN-generated data to enhance rare medical datasets and enhance AI training for CHD diagnosis.

\subsection{Electrocardiogram}

We believe that after echo, ECG is the second authentic signal for CHD diagnosis. Similar to echo, ECG diagnosis necessitates expert knowledge. The ECG identifies irregularities in cardiac rhythm and evaluates the occurrence of prior myocardial infarctions, ischemia, and other conditions that may precipitate heart failure. Due to its inherent safety and cost-effectiveness, ECG is a widely adapted tool for CHD diagnosis \citep{r3}. 

ECG is effective in the diagnosis of both fetal heart structure and functionality. The four-chamber view of the kids heart is of exceptional performance within the realm of parental imaging. Clinicians use and widely accept this perspective because it can serve as a cornerstone of the CHD diagnosis. The visualization with ECG facilitates a thorough analysis of heart architecture and dynamics, allowing clinicians to identify irregularities promptly and formulate tailored treatment regimens. Adult cardiology is very thoroughly explored by ML experts using ECG modality \citep{Ref16,Ref161,Ref162,Ref163,Ref164,Ref17}. In the following paragraphs, we discuss different datasets reported in the literature for CHD recognition using ECG.

\subsubsection{CHDdECG}

An excellent dataset using ECG is proposed by Chen et al. \citep{Chen}. The 93,127 pediatric ECG cases in the dataset came from two hospitals in China: Guangdong Provincial People's Hospital and Shengjing Hospital of China Medical University. The data was collected between 2014 and 2023 using two different ECG machines, the GE MAC800 and the NIHON KOHDEN ECG-2550. Cases include kids who are an average of 2.12 ± 1.50 years old, and CHD cases make up a big part of that group. The information shows real-life medical situations. Most of the CHDs are related to VSD, ASD, PDA, and TOF. As long as the data annotations meet ICD-10 standards, they are clinically relevant. Multi-centered sources and a variety of devices make the dataset more reliable and useful for real-world use.

The dataset underwent extensive preprocessing to ensure excellent signal quality and extract useful features. The authors employed filters to eliminate noise from the raw ECG data and utilized wavelet decomposition to extract the most crucial features. Consequently, they extracted three types of features: waveform features, clinically significant human-concept features, and wavelet-based characteristics. This dataset is open to the public for non-commercial study.

\subsubsection{Physionet}

This dataset \citep{Ref7} is collected from ECG modality. The ECG signals are taken from pregnant women. The database includes data from both typical and unusual women. An extended monitoring time period of 20 hours is used to record ECG signals for Physionet. The Frequency used during the acquistion time is 250 Hz. The research dataset includes two cohorts of pregnant women. Initially a group of 15 women, and then a second group contains 18. WOmen in the first group have babies where CHDs were diagnosed and the second group was healthy.

\subsubsection{Foetuses-CHDs}

A foetuses dataset with 386 non-invasive foetal ECG readings from six medical centres in the Netherlands is presented in the study \citep{Vullings}. The study involved healthy women carrying a single baby, and collected data from foetuses approximately 20 weeks into their development. The dataset is split into two groups: 266 measurements from healthy foetuses and 120 measurements from foetuses who were diagnosed with CHD via US and later proven to have the disease after birth. Some of the heart problems that can occur include atrioventricular and ventricular septal defects. Other heart problems include transposition of the great arteries or ToF, conotruncal defects, and hypoplastic single ventricle defects. The study labels data as either having a CHD or not having one.

The foetal ECG signals were taken with a prototype monitoring system. The signals were then processed to get rid of noise and separate the foetal ECG parts. These were turned into three-dimensional vectorcardiograms, fixed to account for foetal movements. 

\subsection{Computed Tomography Scan}

CT imaging is an essential tool for diagnosing CHD, offering detailed cross-sectional views of the heart and blood vessels. Echo may not fully detect structural abnormalities like septal defects and anomalous vessels, but CT imaging provides precise visualisation of them. Advanced techniques like ECG-gated CT minimise motion artefacts, enhancing diagnostic accuracy for dynamic cardiac assessments. Despite its precision, CT use in CHD diagnosis is carefully balanced against the risks of radiation, particularly in paediatric patients, and is often used as a complementary imaging modality.

\subsubsection{ImageCHD}
The ImageCHD \citep{Xu2} dataset comprises 3D CT images obtained from a Siemens Biograph 64 machine, encompassing 110 patients. The dataset encompasses 16 varieties of CHD. Due to the intricate nature of the structure, a team of four cardiovascular radiologists with substantial expertise in CHD performed the labelling, which included segmentation and classification. A single radiologist assigns the segmentation label to each image, while four radiologists conduct its diagnosis. The average duration of labelling each image is approximately 1 to 1.5 hours.

\subsection{Non-clinical Dataset}
Non-clinical data can enhance CHD recognition through ML by offering additional insights beyond conventional clinical measurements. Socioeconomic status, parental health behaviours, environmental exposures, healthcare access, and demographic data can be incorporated into ML models to improve predictive accuracy. Maternal exposure to contaminants during gestation or dietary inadequacies may serve as significant predictors of CHD risk. By integrating non-clinical characteristics, ML algorithms can reveal concealed patterns and correlations that may not be evident from clinical data alone, providing a more comprehensive and precise risk assessment for early identification and management. Three such datasets are reported in the litrature which we present in the following pargraphs.

\subsection{Guangzhou Hospital Dataset}

This dataset \citep{Liang} comprises electronic health records from 567,498 pediatric patients at the Guangzhou Women and Children’s Medical Centre. The records encompass comprehensive clinical data, including diagnoses classified per the ICD-10, symptoms, history of present illness, physical examination results, and laboratory test outcomes. The dataset encompasses a varied paediatric population with a median age of 2.35 years and includes a broad spectrum of both common and unusual paediatric illnesses. All patient records were anonymised to guarantee adherence to data privacy standards. 

\begin{figure*}[t]
	\centering
	\includegraphics[width=10 cm, height=7cm]{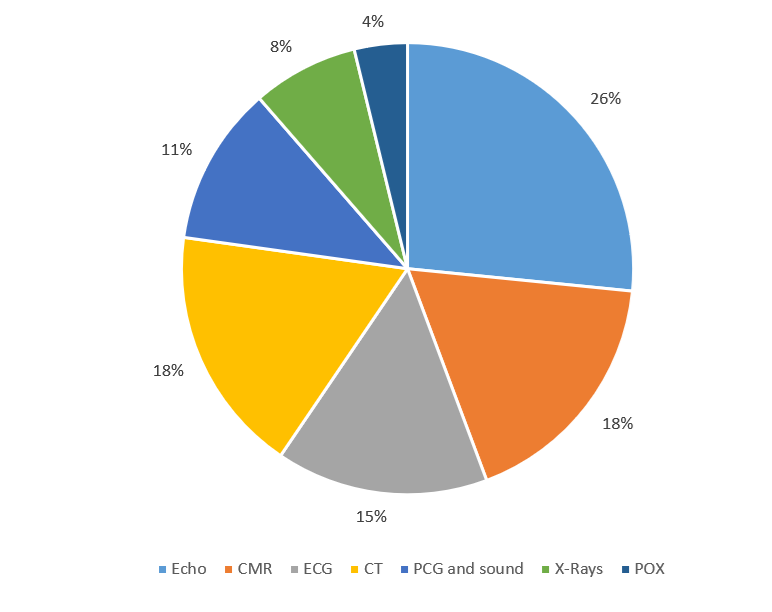}
	\caption{Research papers reported so far for each modality using CHD-ML.} 
	\vspace{0,25cm}
	\label{Fig-Modalities}
\end{figure*}

\subsubsection{Shanxi Province Dataset}

This dataset \citep{Luo2} comprises records of 33,831 live births collected between 2006 and 2008 in Shanxi Province, China. Among these, 78 individuals were identified with CHD abnormalities. The dataset exhibits significant imbalance, with the majority of entries classified under the non-CHD category. The research sought to forecast CHD risk by examining maternal health and lifestyle determinants, encompassing characteristics such as maternal age, income, familial history, medical history, nutrition, medication consumption during gestation, and environmental exposures (e.g., X-rays). The information utilized in this analysis originates from a comprehensive, retrospective, population-based epidemiological survey executed by the Population and Family Planning Commission of Shanxi Province, China. During preprocessing, a ``total risk factor score" was calculated to reduce the number of dimensions and fix any class imbalances. 

\subsubsection{University of California Dataset } 
The UCI CHD dataset is sourced from the publicly accessible repository of the UCI \citep{Ref43}, available on the Kaggle website \footnote{https://www.kaggle.com/datasets/redwankarimsony/heart-diseasedata}. The dataset acquired included information from 1,050 patients. Out of 76 overall attributes, only 14 were utilised in predicting heart disease. This is due to the fact that the other traits exert a lesser influence on the disease relative to these attributes. Before categorisation, the dataset is subjected to a cleaning and filtering process to eliminate missing or redundant information.

\subsection{Summary}
CHD recognition relies on a range of diagnostic modalities tailored to identify structural abnormalities and assess heart function. In this part of the paper we listed and discussed different kinds of modalities that can be used for CHD recognition. We must like to add here that there might be some other methods (sub-methods) which can be used for CHD recognition. We discussed only those methods which are used for CHD recognition combined with ML.

The figure \ref{Fig-Modalities} delineates the contributions of the above mentioned modalities in research papers pertaining to CHD recognition. Diverse methods have been employed in research publications for the recognition of CHD, each contributing differently. Echo is the most utilised modality at 26\%, CMR and ECG each account for 18\%, highlighting their importance in detecting structural and electrical anomalies, respectively. CT constitutes 15\%, PCG and acoustic methodologies include 11\%, X-rays account for 8\% and are typically utilised for general anatomical assessments, whereas POX comprises 4\%.

\section{ML/DL Algorithms Used For CHD}
\label{sec-5}
It is extremely challenging to organize all ML methodologies for CHD recognition into a single taxonomy \citep{H1,H2,H3}. The difficulty resides not in the plethora of techniques suggested by ML specialists but in the variety of data formats employed for problem analysis. The methodologies employed to assess ECG signals are distinct from those utilised for the examination of sound modalities. Therefore, we do not adhere to a specific taxonomy to address CHD through ML. We will strive to link each model with its respective implementation. In the following paragraphs, we provide a thorough analysis of these strategies and a review of the academic works that specifically address them. Furthermore, we provide a comprehensive detail of the benefits and drawbacks linked to each strategy.

Table \ref{yearwise} clearly indicates that, throughout the years, researchers have shown greater interest in DL algorithms rather than classical ML in the development of CHD-ML models. Notably, 60 out of 74 papers (81\%) focused on addressing CHD using DL, while only 14 papers (19\%) utilized classical ML methods.

\subsection{Conventional ML Methods}

Machine learning has shown extemely good performance in various compute vision tasks, including, NLP\citep{far1}, face analysis, medical diagonostics etc. \citep{C1,C2,C3,C4,C5,far2,far3,far4,far5}. Different TML techniques have been used for CHD recognition. Feature engineering for classical ML algorithms is a more complex and labor-intensive procedure. TML consists of multiple distinct phases, namely pre-processing, feature extraction, and classification.

The work proposed in \citep{Liu2} used a dataset collected from Shanxi Province, China. Nine predictor variables related to maternal and environmental risk factors were derived for analysis. Three classification models, SVM, RF, and LR were applied to address the dataset's imbalance. The models were evaluated using metrics like TPR, TNR, $A_c$, G-Mean, and AUC. SVM performed the best, achieving a TPR of 69.23\%, TNR of 94.76\%, $A_c$ of 94.70\%, and AUC of 81.87\%, making it most effective in identifying high-risk groups while maintaining strong specificity.

The research conducted by Xu et al. \citep{Ref8} employs cardiac signals to classify CHD. The authors have identified a variety of diverse and extensive characteristics. The authors employed frequency domain data and wavelets to construct traditional classifiers within an RF framework. The authors of the research contend that SOA datasets have markedly enhanced outcomes. The researchers replicated the anatomical architecture of the hearts and principal blood arteries of 29 youngsters with CHD anomalies. The authors reported better results compared to previous findings.

The work in \citep{Ref12} leverages the ZCHSound database, an open-source pediatric CHD heart sound dataset. Classification was performed using RF and KNN. Evaluation criteria included $A_c$, $S_c$, $S_p$, and $F_c$. The $F_c$ attained for high-quality data was 0.90 for binary classification (normal vs. CHD) and 0.934 for multi-class classification (normal, ASD, VSD, PDA, PFO). The F1-score for low-quality data was 0.62 for binary classification and 0.46 for multi-class classification. Some more papers that are using SVM, RF, LR, GB, and Adaboost classifiers with certain features can be explored in \citep{Siefkes,Ref26,Truong,Xu1,Thomas}.

The manuscript \citep{Ref19} presents an approach for heart sound segmentation utilising an HSMM with variable sojourn time parameters. The HSMM analyses PCG signals by utilising aspects including homomorphic and Hilbert envelograms, wavelet-based characteristics, and power spectral density. The datasets used during this work is PhysioNet. The evaluation metrics include $A_c$, $S_c$, $S_p$, and $F_c$. The authors report an F1-score of 92\% on the PhysioNet dataset. The method is resilient across datasets, facilitating efficient transfer learning across training and testing sets.

TML employs NNs for classification tasks, discerning correlations within data through recurrent training processes. They play a significant role in image recognition and classification, identifying patterns such as forms, edges, or textures within the data. The research reported in \citep{Lili} employs a BP NN as the ML technique for identifying CHD from cardiac sound data. Features are retrieved using two methodologies. The dataset consists of 200 heart sound recordings, evenly divided between normal and pathological instances. The model attained a $S_c$ of 0.88 and a $S_p$ 0.93.

Another work reported in \citep{Ref59} employs an ANN as the principal ML technique for CHD recognition in prenatal contexts, utilizing non-clinical data from expectant moms. The dataset comprises 33,831 live birth records gathered from Shanxi, China. The ANN attained an exceptional $A_c$ of 99\%. The authors improved  $A_c$ of their previous own work, (previously using SVM with an $A_c$ of 0.947). Some more excellent papers that are exploring ANNs for CHD recognition can be studied here \citep{Lv1, Belinha1,He, Sakai}.

\subsection{Deep Learning Based Methods}

This section summarizes various DL models developed for CHD recognition. The progression of DL algorithms has significantly transformed CV methodologies. Historically, CV predominantly relied on manually designed features and techniques to analyse visual data, occasionally resulting in suboptimal performance in intricate jobs. Nonetheless, the application of DL, especially CNNs, has fundamentally transformed this area of study. DL models may autonomously obtain hierarchical representations from raw data, thereby enhancing their capacity to accurately identify complex patterns and features. This innovative technology has markedly improved the accuracy, resilience, and scalability of CV systems in multiple areas, including object detection, image classification, facial recognition, and medical imaging. Table \ref{yearwise} indicates that the predominant algorithms used in contemporary research are based on DL methods. We present a summary of the most commonly used DL strategies for CHD recognition in this section of the paper.

\subsubsection{Convolutional Neural Networks}

The architecture of CNNs is modelled inspired from human visual system \citep{169,170}. The CNNs framework has two phases: feature extraction and classification. The feature extraction consists of multiple hierarchical structures within convolutional, activation, and lastly pooling layers. The convolutional layer generates features that are based on input data through the convolution process. The model's complexities are reduced, enabling a more simple training procedure due to the reduced number of hyperparameters requiring adjustment through the spatial sharing of the kernel throughout the entire input data in each convolutional layer. A pooling layer reduced the resolution of the feature map, hence achieving shift invariance. The pooling layer is situated subsequent to each convolutional layer. In the classification step, the feature maps generated during feature extraction are utilised with a SoftMax function and fully connected layers. The authors of \citep{174} shown that the fully linked layer can be eliminated by employing a global average pooling layer. SoftMax is frequently employed for classification tasks \citep{175}, \citep{176}, while SVM and other classifiers have also been used. Several exemplary research studies that are using CNNs for CHD recognition are discussed below.

\begin{itemize}
    \item A research paper that leverages time-frequency representations of PCG information and CNNs is reported in \citep{Bozkurt}. The principal feature extraction technique included in the study is sub-band envelopes, an innovative method in this field, which has demonstrated superior performance compared to traditional features like Mel Frequency Cepstral Coefficients and Mel-spectrograms. The analyzed datasets consist of heart murmur signals, which contain pediatric PCG recordings with clinical annotations. Another publicly available dataset, PhysioNet, has also been used. The evaluation measures reveal better performance, with $S_c$ 0.845, $S_p$ 0.785, and $A_c$ 0.815.

    \item Similarly, \citep{Vullings} uses deep NNs that include six convolutional layers and residual connections to examine foetal ECG data for CHD recognition. Feature extraction entailed the generation of a three-dimensional vectorcardiogram from fetal ECG signals obtained through maternal abdominal electrodes. The dataset consisted of 386 foetal ECG recordings, encompassing 266 from healthy foetuses and 120 from those with CHD. The proposed method attained an $A_c$ value of 0.76.

    \item A  CNN-based method for CHD is proposed in \citep{Ref13}. The authors present a method named CHDdECG. The type of data used is ECG signal. Along with DL wavelet transform is also used. The authors subsequently amalgamate these characteristics with substantial human-concept variables. CHDdECG was assessed on a dataset of 65,869 samles obtaining a good AUC value. The proposed method was assessed using two separate external datasets, comprising 7137 and 8121 cases.

    \item Kavitha et al. \citep{Ref12} proposed a method termed multilayer deep detection perception. The authors of the paper employed ultrasonic imagery for their investigation. The model derives features using a multilayer DL architecture with several perceptron layers. The authors assert that the suggested model's performance on the SOA dataset is superior.
    
    \end{itemize}

\subsubsection{GAN Models}

A GAN is a category of ML frameworks intended to produce novel data samples that mimic a specified dataset. GANs comprise two neural networks: the generator, responsible for producing synthetic data, and the discriminator, tasked with assessing the veracity of the input. The generator seeks to refine its proficiency in producing realistic data, while the discriminator augments its capacity to differentiate between authentic and fake data. Unlike CNNs, which primarily focus on image classification, object identification, and feature extraction, GANs are specifically designed for data production. We believe that GANs will be exceptionally proficient at CHD recognition, particularly in contexts characterised by restricted data availability. The rarity of the ailment and the challenges of obtaining labeled medical data frequently limit the size and unevenness of CHD datasets. GANs can produce realistic synthetic data, including echos and X-rays, to enhance the dataset and optimize model training. GANs also mitigate class discrepancies by generating more samples for under-represented circumstances, hence improving classification performance. Furthermore, GANs discern intricate patterns in medical photos, essential for detecting difficult CHD situations. The capacity to generate high-quality, diversified data renders GANs an essential instrument for enhancing CHD recognition in resource-limited environments.

\begin{itemize}
    
\item A paper in \citep{gong197} proposes the DANomaly and GAN-based model for fetal CHD recognition. Expert cardiologists annotate the end-systolic phases of four-chamber heart video slices to extract features. Evaluation metrics reveal that the method achieves 85\% recognition accuracy, outperforming expert cardiologists and SOA networks.

\item A work proposed in \citep{Ref58} utilises GAN to produce synthetic cardiac MRI images for patients with CHD, particularly those diagnosed with TOF. The GAN was trained using 303 MRI datasets from patients, generating over 100,000 synthetic images. The synthetic images were utilised to train U-Net segmentation models for the segmentation of heart chambers. The evaluation measures comprised the Dice coefficient and percentage area variation. U-Nets trained on synthetic data had comparable performance to those trained on authentic patient data, with a segmentation accuracy discrepancy of less than 1\%, so illustrating the viability of utilising synthetic datasets for training purposes.

\end{itemize}

\subsubsection{Recurrent Neural Networks}

RNNs and LSTMs are a kind of NNs engineered to capture long-range relationships in sequential data through the use of memory cells \citep{Hochreiter}. LSTMs are predominantly used for time series and NLP applications, but they can facilitate image recognition by examining sequences of image data, including video frames or segmented image characteristics \citep{Donahue}. LSTMs improve tasks like action detection in movies and image sequence prediction by maintaining contextual information; hence, they augment conventional spatial analysis techniques in image-centric applications \citep{Srivastava}.

\begin{itemize}
  
    \item The research paper in \citep{Liu2} utilises LSTM for the classification of CHD based on ECG signals. Symbolic aggregate approximation does feature extraction, which reduces the number of dimensions in the data and turns sequences into symbolic representations while keeping important characteristics and lowering the amount of work that needs to be done on the computer. The dataset comprises 5,000 preprocessed heartbeats annotated with class values from automatic annotations. The model achieves an $A_c$ of 98.4\%. Symbolic aggregate approximation preprocessing markedly improves classification performance and efficiency, yielding optimal results.

    \item An LSTM based framework has been introduced by Ng et al. \citep{Ref11} for CHD classification. The authors of this study have presented a novel and unique feature extraction method. This study integrates both chromatic and textural attributes. Better results are reported by this paper as compare to previous results on their own dat.

    \item A recent study on CHD recognition utilising DL is presented in \citep{Ref7}. This paper introduces an ML-based technique for CHD recognition, termed ML-CHDPM. This method focuses on combining LSTM with specific attention mechanisms. This study presents an approach that integrates CNN, bi-directional LSTM, and attention mechanisms. We have observed improvements in the results with this method. The technique utilises ECG signals acquired from pregnant women. The ML-CHDPM is trained on an extensive dataset comprising 33 candidates. The initial 15 moms have a foetus with abnormalities, but the subsequent 18 have normal foetuses. The initial 15 patients are once more categorised into CHD and severe CHD groups. The suggested approach identifies significant patterns and correlations in the data, leading to accurate classification. The model has been assessed using $S_c$, $S_p$, and $A_c$ . The authors claim that the new model has enhanced previously published results.

\end{itemize}

\subsubsection{Hybrid Models}
Hybrid models integrate TML methods with DL to capitalise on the advantages of both paradigms. These models frequently employ conventional ML techniques, such as DT or SVM etc., to analyse structured or tabular data, whereas DL models, such as neural networks, manage unstructured data, like images or text \citep{Dong}. In image recognition tasks, a DL model can extract high-level features from images, whereas a conventional ML method may classify these features according to established criteria \citep{Razzak}. This combination enhances performance by integrating DL's ability to process complex data with the interpretability and efficiency of TML methods, making hybrid models versatile in applications such as medical diagnostics.

\begin{itemize}
    \item A paper presented in \citep{Liang} blends a hierarchical LR classifier with a DL-based NLP system to extract clinically important features from pediatric electronic health records. The dataset comprises 1,362,559 outpatient visits from 567,498 paediatric patients. The model achieved a median $A_c$ of 0.90 across several diagnostic levels. Compared to prior works employing comparable hierarchical methodologies, this study leverages a substantially bigger dataset and includes harmonised data inputs for increased diagnosis accuracy and robustness, particularly for complicated or unusual illnesses.

    \item This study \citep{Ref15} employs a hybrid ML method combining CNNs, BiLSTM for the early detection and diagnosis of CHD using ECG signals. The dataset includes ECG records sourced from the PhysioBank repository and other publicly available databases, with preprocessing steps involving Z-score normalization and segmentation into 2-second intervals. The proposed ML-CHDPM model achieves superior evaluation metrics, with an $A_c$ of 0.96, $P_c$ of 0.89, $R_c$ of 0.99, and $S_p$ of 0.90.
    
     \item A paper in \citep{ref6} details the development of an AI-based CHD model (CHDNet). This model works as a binary classifier that analyses echo to identify abnormalities in the heart. The authors claim that CHDNet's effectiveness is equal or better than that of medical specialists. The authors propose two techniques: Bayesian inference and dynamic neural feedback. These methods are employed to precisely evaluate and improve the diagnostic reliability of AI. The preliminary method enables the neural network to generate an evaluation of its reliability rather than a single forecast outcome. The last method uses a computational neural feedforward cell, enabling the neural network to convey information from the output layer to the superficial layers. This allows the neural network to selectively activate specific neurones. The researchers evaluated the effectiveness of the two techniques by training on 4,151 echo signals. The trained CHDNets was evaluated using an internal test set of 1,037 echo and an additional set of 692 videos obtained from several cardiovascular imaging systems. Each echo film is linked with an individual. This research illustrates the impact of Bayesian inference on different neural network architectures and measures the notable performance difference between internal and external test datasets.

\end{itemize}

\subsubsection{Deep Residual Networks}

A ResNet is a deep NN architecture developed to mitigate problems such as vanishing gradients and performance decline in extremely deep networks \citep{He}. It employs residual learning, utilising shortcut connections that enable the model to circumvent specific layers by simply summing the input with the output of a layer. This facilitates the effective training of extensive networks and enhances precision \citep{Huang}.
\begin{itemize}
    \item Pachiyannan et al. \citep{Ref15} introduced a system based on residual learning, termed RLDS. The approach utilizes a residual learning mechanism specifically intended for fetal CHD. The RLDS derives unique characteristics from echo data. The RLDS produces attention maps that allocate significance scores to each feature. This facilitates the comprehension of the diagnosis procedure for medical practitioners. The authors indicated improved accuracy on the SOA dataset.

    \item ResNet18 and ResNet50 are different versions of the ResNet architecture, developed to mitigate vanishing gradients and enhance performance in deep networks using residual learning. ResNet18 comprises 18 layers featuring fundamental residual blocks, rendering it computationally economical and appropriate for smaller datasets or resource-limited settings. Similarly, ResNet50 comprises 50 layers with bottleneck blocks, enabling it to discern more intricate patterns, thereby rendering it suitable for extensive datasets and precision-oriented jobs. ResNet18 is more efficient and requires fewer resources, while ResNet50 provides superior accuracy but entails greater computational and memory requirements, making it adaptable for various image recognition applications. Researchers use ResNet50 and ResNet 18 for CHD recognition. Papers can be explored for corresponding methods in \citep{Ref15,Ref13,Han} and \citep{ref6,Xuu}.

    \end{itemize}

\subsubsection{Models for Non-clinical Data} 

The study conducted by Cheng et al. \citep{Ref17} involves analysing the 2D and Doppler transthoracic echo of children from two distinct clinical cohorts at BCH. A DL framework was created to identify cardiac views, integrate data from many perspectives and modalities, visualise high-risk areas, and assess the likelihood of an individual being healthy or having an ASD or a VSD. The DL models that integrate many modalities and scanning perspectives show enhanced performance. Integrating diverse perspectives and methodologies of transthoracic echo into the model enables the precise recognition of infants with CHD without necessitating invasive treatments. This indicates the possibility of enhancing CHD detection efficacy and streamlining the screening procedure.

CHD can occur in diverse forms and exhibit subtle signs that may be evident from birth. The research study referenced in \citep{Ref15} introduces an innovative healthcare application known as the ML-based CHD Prediction Method named ML-CHDPM. This strategy aims to address challenges and enhance the precise diagnosis and classification of CHD in pregnant women. The ML-CHDPM model uses advanced ML techniques to categories instances of CHD. The method utilizes relevant clinical and demographic variables. The model has been trained on a comprehensive dataset, allowing it to discern nuanced patterns and connections.

\subsubsection{You Only Look Once for CHD}

Some papers also explore YOLO models for CHD recognition. YOLO is a real-time object recognition framework that swiftly and accurately identifies and localises several items inside an image \citep{Redmon}. In contrast to conventional techniques that use sliding windows or region proposals for image scanning, YOLO approaches the detection challenge as a unified regression task. It partitions an image into a grid and concurrently predicts bounding boxes, confidence scores, and class probabilities for each grid cell, rendering it very rapid and appropriate for real-time applications \citep{Redmon18}. YOLO's efficacy and straightforwardness have rendered it extensively utilized in domains such as video surveillance, autonomous driving, and robots \citep{Bochkovskiy}.

YOLO has undergone substantial advancements from YOLOv4 to YOLOv10, enhancing the speed, precision, and diversity of object recognition. YOLOv8 used transformer-based attention to enhance small object identification, whereas YOLOv9 focused on lightweight architectures for IoT and mobile devices. YOLOv10 featured modular architectures and visual transformers for enhanced precision and scalability, facilitating applications in autonomous systems and medical diagnostics. A paper proposed in \citep{Zhixin} uses YOLO7 for CHD recognition. Similarly, YOLO5 has been used for CHD recognition in the paper \citep{Sakai}. Another paper that was proposed in 2021 is using YOLO4 for CHD diagnosis \citep{Gearhart}.

\subsubsection{U-Net Architectures}

U-Net models are highly efficient CNN designs tailored for image segmentation. Their U-shaped architecture has an encoder that stores contextual information and a decoder that helps with accurate localisation. Spatial features are kept by skip connections \citep{Ronneberger}. This design is especially good for detecting CHD because it makes it easier to separate out complex heart structures from medical images like echo, MRIs, or CT scans, which is important for diagnosis and planning therapy \citep{cicek}.

\begin{itemize}

    \item The study reported in \citep{Cheng} utilises an ensemble of four 3D U-Net models for whole-heart segmentation from CMR. The 3D U-Nets were trained using a dataset including 20 manually segmented pictures, utilising data augmentation methods such as affine transformations and Gaussian noise to improve generalisation. The architecture had four tiers with ascending feature channels and max-pooling for down-sampling. A categorical cross-entropy loss with weights that changed in space fixed class imbalances and improved the accuracy of the borders. The ensemble outputs were averaged to generate final segmentation maps, resulting in effective segmentation despite anatomical heterogeneity and constrained training data.

    \item Khan et al. \citep{Ref75} introduced a DL model called the CDLM. TO delineate four chambers of heart, authors use a segmentation framework. The authors employ a graph matching technique to extract link data and categorize all the boats. A  dataset of 68 CT scans images has been used in this work. The authors assert that the new technique produced better results than those previously published. Similarly, work reported in \citep{Ref12,Wang} also uses 3D U-Net models for CHD recognition.

\end{itemize}

\subsection{Summary}
While DL techniques are well-established in the broader field of ML applications, their adoption in the domain of CHD recognition is relatively recent. Despite their promising capabilities, the application of DL methods for CHD identification has been sporadic and is yet to be fully explored. There remains significant potential to uncover the full utility of DL in this context. This study focuses on recent advancements in ML techniques, particularly over the past eight years, with an emphasis on their role in CHD recognition. By analyzing these developments, we aim to shed light on emerging trends and highlight opportunities for further research and innovation in this critical field.

\captionsetup[longtable][t]{skip=0.5cm} 
\setlength{\tabcolsep}{4pt} 
\renewcommand{\arraystretch}{1.2} 
\begin{longtable}{p{3.5cm}p{2.5cm}p{2cm}p{3.5cm}p{4cm}}
\caption{CHD Databases Reported in State-of-the-Art (SOA).}
\label{yearwise}\\
\toprule
\textbf{Authors} & \textbf{ML Method} & \textbf{Data Type} & \textbf{Dataset Used} & \textbf{Evaluation Metrics} \\ 
\midrule
\endfirsthead

\toprule
\textbf{Authors} & \textbf{ML Method} & \textbf{Data Type} & \textbf{Dataset Used} & \textbf{Evaluation Metrics} \\ 
\midrule
\endhead

\midrule
\multicolumn{5}{r}{\textit{Continued on the next page...}} \\
\bottomrule
\endfoot

\bottomrule
\endlastfoot

\multicolumn{5}{l}{\textbf{2024}} \\ \midrule
Jia et al. \citep{Ref12} & RF, KNN & PCG & ZCHSound & $F_c$: 0.90 and 0.934\\
Qiao et al. \citep{Ref15} & Hybrid & ECG & PhysioBank & $A_c$: 0.96, $P_c$: 0.89, $R_c$: 0.99, $S_p$: 0.90 \\
Cheng et al. \citep{Ref13} & CNN & ECG & CHDdECG &   RUC: 0.915, $S_c$: 0.917, $S_p$: 0.90 \\
Zhixin et al. \citep{Zhixin} & 3D U-Net, 2D U-Net, and BiConv LSTM	& CT &	3255 frontal preoperative chest radiographs & $S_c$: 0.82, $S_p$: 0.87, and $A_c$: 0.86,  $DS_c$: 0.73\%\\
Han et al. \citep{Han}	& ResNet18	&	X-Rays & DICOM X-rays & $A_c$: 0.80 , ROC: 0.85\\
Xuu et al. \citep{Xuu}	& ResNet18, DenseNet121, MobileNetv2	& Echo, CT, and MRI & More than 3750 images & AUC: 0.94, $S_c$: 0.97, $S_p$: 0.98\\
Nguyen et al. \citep{Nguyen}	& 3D U-Net and BiConv LSTM &	CT &	139 CT scanns collected at Guangdong  Hospital
& $S_c$: 0.82, $S_p$: 0.87, $A_c$: 0.86, $DS_c$: 0.73 \\
Cheng et al. \citep{Cheng} & ResNet18	& Echo & 1932 children
&	AUC: 0.99, $A_c$: 0.99 \\
Marelli et al. \citep{Marelli}	& GBDT, SVM	& non-clinical data		& 19,187 patients, with 3784 labeled as true CHD cases & $S_c$: 0.99, $S_p$: 0.97,  $F_c$: 0.99 \\
Pace2 et al. \citep{Pace2}	& 3D U-Net & CMR & HVSMR-2.0 & $DS_c$: 0.87\\
Siefkes et al. \citep{Siefkes} & RF, LR, GB, saturation-level & POX &	$A_c$: 0.92 \\ \\ 

\multicolumn{5}{l}{\textbf{2023}} \\ \midrule
Xuu et al. \citep{Xuu}	& CDLM & non-clinical data & Over 3,750 CHD patients & $S_c$: 0.91, $S_p$: 0.92, PPV: 0.90, NPV: 0.55\\
Pachiyannan et al. \citep{Ref15}	& ANN &	Echo	&	CHD-HSY & $A_c$: 0.71, $S_c$: 0.63, $S_p$: 0.77 \\
Sapitri et al. \citep{Sapitri}	& YOLOv7	& Echo	& 40 fetal echocardiography videos & $P_c$: 0.82 \\
Arnaout et al. \citep{Arnaout}  & ResNet50	& Echo	& Seven-CHDs & AUC: 0.92 $A_c$: 0.91\\
Yang et al. \citep{Yang}	& ResNet50 and LSTM &	Echo	& Seven-CHDs	& $A_c$: 0.94, $S_c$: 0.93, $S_p$: 0.95, AUC: 0.96\\
Jia et al. \citep{Ref12}  & CNNs	& Echo	& &	AUC: 0.99, $S_c$: 0.95, $S_p$: 0.96 \\
Jiang et al. \citep{Jiang}	& CNNs	& US & 9,822 images from 125 fetal echocardiograms &	AUC: 0.99, $S_c$: 0.95, $S_p$: 0.96\\
Rima et al. \citep{Arnaout}	& RNN &	Heart murmurs	& CHD-HSY &	$A_c$: 0.97\\
Yang et al. \citep{Yang} & Hybrid	& CT & Guangdong  Hospital &	$S_c$: 0.82, $S_p$: 0.87, $A_c$: 0.86 \\
Tan et al. \citep{ref6}   & Bayesian inference models & Echo & -- & $P_c$: 0.94, $R_c$: 0.95\\
Xu et al. \citep{Xuu} & 3D U-Net, 2D U-Net, and BiConv LSTM &  & ROC: 0.79, $A_c$: 0.69\\ 

\multicolumn{5}{l}{\textbf{2022}} \\ \midrule
Qio	& RLDS (CNNs) & US &	2,100 fetal Four-Chamber views
 &	$P_c$: 0.93, $R_c$: 0.93, $F_c$: 0.93, $A_c$: 0.93 \\
Gearhart et al. \citep{Gearhart}	& CNNs & Echo & Boston Children’s Hospital: 18,264, Leukemia Patients: 32,759
& $A_c$: 0.90, PPV: 0.98, $S_c$: 0.94\\
Truong et al. \citep{Truong} 	&RF&	Echo & fetal Echo derived from 3910 singleton fetuses
&	$S_c$: 0.85, $S_p$: 0.88, PPV: 0.55, NPV: 0.97\\
Goretti et al. \citep{Goretti}	&C NNs&	non-clinical& Santa Maria Nuova Hospital, Florence, Italy with 1,888 entries
&		ROC: 0.79\\
Liu1 et al. \citep{Liu1}	& Residual CNNs &	PCG	& Total recordings: 884 heart sound recordings
& $S_c$: 0.93, $S_p$: 0.99, $P_c$: 0.88, $A_c$: 0.94\\
Xu et al. \citep{Xu1} 	&RF and AdaBoost&	PCG& Total recordings: 941 pediatric heart sound recordings
&	$A_c$: 0.95, $S_c$: 0.94, $S_p$: 0.96, $F_c$: 0.95\\
Oliveira et al. \citep{Oliveira}  &	CNNs &	PCG&	CDD &	$S_c$: 0.94, $S_p$: 0.95, $F_c$: 0.94\\
Nurmaini et al. \citep{Nurmaini}  &	DenseNet201 and CNNs &	Echo & Data collected from 76 pregnant women during routine screening
&	$S_c$: 0.91, $S_p$: 0.92, $A_c$: 0.92\\
Kavitha et al. \citep{Kavitha}	& multilayer deep detection perceptron &	US	& fetal cardiac ultrasound screening of 160 cases
&	$A_c$: 0.98\\
Sakai et al. \citep{Sakai} &	deep neural network (autoencoder) & US	&	ROC: 0.97\\
Wu et al. \citep{Wu}	& YOLOv5 & US	& Total 1,695 fetal ultrasound images.
& $P_c$: 0.96, $R_c$: 0.85\\ 

\multicolumn{5}{l}{\textbf{2021}} \\ \midrule
Komatsu et al. \citep{Komatsu}	& CNN	& US	& Total 363 pregnant women
&	ROC: 0.78 \\ 
Xu et al. \citep{Xu2}	& 3D-CNNs	& CT	& ImageCHD	& $A_c$: 0.81\\
Wang et al. \citep{Wang} 	& CNNs & 	Echo	&Total Patients: 1,308 children
 & 	$A_c$: 0.93 (binary classification), 0.92 (three-class classification)\\
Lv et al. \citep{Lv1}	& SVM &	PCG	&	Heart sound recordings from 1,397 pediatric patients
& $S_p$: 0.97, $S_c$: 0.89, $A_c$: 0.96\\
Belinha et al. \citep{Belinha1} & ANN and DT & PCG	&1,655 individuals included after applying exclusion criteria
&	AUC: 0.76\\
He et al. \citep{He}	& CNN	& US &	& $DS_c$: 0.98\\

\multicolumn{5}{l}{\textbf{2020}} \\ \midrule
Tan et al. \citep{Tan} & SVM & Echo & DICOM videos of 100 patients & $F_c$: (0.72-0.87) \\
Qiao et al. \citep{Qiao}	& YOLOv4 &	Echo	& 1,250 Four Chamber ultrasound images
&	$P_c$: 0.91, $R_c$: 0.97, $F_c$: 0.94\\
Gong et al. \citep{Gong}	& GANs	& Echo	&	Total 3596 images & $A_c$: 0.85\\
Karimi et al. \citep{Karimi}	& GANs	& CMR	& 64 CMR (2-18 years) & $DS_c$: 0.91 and 0.84\\
Thomas et al. \citep{Thomas}	& SVM   & PCG	&   PhysioNet	& $A_c$: 0.85\\
Diller et al. \citep{Ref58} & PG-GAN,U-Net &  MRI & -- &   $DS_c$: 1  \\
Rani et al. \citep{Ref59} & ANN & non-clinical data & Shanxi CHD &   $A_c$: 0.99\\

\multicolumn{5}{l}{\textbf{2019}} \\ \midrule
Oliveira et al. \citep{Ref19} & HSMM & PCG & PhysioNet & $A_c$: 0.92\\
Bozkurt et al. \citep{Bozkurt} & CNNs & PCG & PhysioNet & $S_c$: 0.84, $S_p$: 0.78, and $A_c$: 0.81\\
Lili et al. \citep{Lili} & NN & PCG & -- & $S_c$: 0.88, $S_p$: 0.93\\
Linag \citep{Liang}  & CNN+LR & non-clinical data & -- & $A_c$: 0.90\\

\multicolumn{5}{l}{\textbf{2018}} \\ \midrule
Luo et al. \citep{Luo2} & SVM, RF, and LR & PCG and heart murmurs &  ZCHSound &  TPR: 0.69, TNR: 0.94, $A_c$:0.94, and AUC: 0.81 \\
Liu et al. \citep{Liu2} & LSTM and RNN & ECG &  Physionet &   $A_c$:0.98 \\
Vullings et al. \citep{Vullings} & CNNs & ECG & NA & $A_c$: 0.76\\ 
\end{longtable}

\section{Discussion and Challenges}
\label{sec-6}
The application of ML techniques in pediatric cardiology offers transformative potential but also faces significant challenges. These AI-driven methodologies hold promise across various domains, including examination and clinical diagnosis, fetal cardiology image processing, risk classification and prognosis, cardiac intervention planning, and precision cardiology. ML algorithms demonstrate substantial potential in diagnosing both critical and non-critical CHDs. However, our findings suggest that significant efforts are still required in both the ML and healthcare sectors to overcome current limitations. Despite advancements, ML models are not yet at the stage where they can reliably and accurately identify complex CHDs independently.

\subsection{Evaluation metrics CHD-ML}

From Table \ref{yearwise} it can be seen that various parameters are used for evaluation of a CHD-ML based model. We define these terminologies for readers in this subsection of the paper. Four terms underpin these measurements: The number of accurately identified infected samples is referred to as TP. TN denotes accurately detected harmful data samples. Likewise, we denote FP as occurrences in which the model erroneously classify healthy samples as infected. A FN refers to the erroneous categorisation of contaminated samples as healthy. Some evaluation measures which are derived from the above mentioned terminologies are presented in Table \ref{equations}. \\

 




\begin{table}[h!]
\centering
\begin{tabular}{|p{5cm}|p{5cm}|}
\hline
\textbf{Metrics} & \textbf{Equation} \\ \hline
Accuracy        & $\text{$A_c$} = \frac{TP + TN}{TP + TN + FP + FN}$        \\ \hline
Precision        &  $\text{$P_c$} = \frac{\text{TP}}{\text{TP} + \text{FP}}$          \\ \hline
Sensitivty        & $\text{$S_c$} = \frac{\text{TP}}{\text{TP} + \text{FN}}$ \\ \hline
Specifity       & $ \text{$S_p$} = \frac{\text{TN}}{\text{TN} + \text{FP}}$ \\ \hline
F1-measure & $  \text{$F_c$} = 2 \times \frac{\text{$P_c$} \times \text{$R_c$}}{\text{$P_c$} + \text{$R_c$}} $\\ \hline
\end{tabular}
\caption{CHD-ML evaluation measures.}
\label{equations}
\end{table}

Along with the above evaluation metric, $DS_c$ is also used for evaliation of CHD-ML modle. $DS_c$ is normally used for segmentation tasks in litrature. The $DS_c$ is a  measure that quantify the similarity between two or more sets. It measures the intersection between the anticipated segmentation and the actual ground truth, on a scale from 0 (no intersection) to 1 (complete intersection).

    \[
    \text{$DS_c$} = \frac{2 |A \cap B|}{|A| + |B|}
    \]

    Where:
\begin{itemize}
    \item \( A \): A set of pixels in the estimated data.
    \item \( B \): A set of pixels in the ground truth data.
    \item \( |A \cap B| \): The number of pixels common to both \( A \) and \( B \).
    \item \( |A| \) and \( |B| \): The total pixels in \( A \) and \( B \), respectively.
\end{itemize}

For binary segmentation masks, it can also be expressed as:

\[
\text{$DS_c$} = \frac{2 \cdot TP}{2 \cdot TP + FP + FN}
\]

\subsection{Challenges to CHD-ML}

\subsubsection{Problems with Data Acquisition} 

The healthcare sector is currently consolidating data from many facilities and patients to identify CHDs. By effectively employing this data, physicians can anticipate more sophisticated treatment options and improve the overall healthcare delivery system for this disease. The duration of data collection for CHD is very limited. Data collection is crucial immediately after delivery. As time progresses, the severity of the illness increases. The majority of individuals assess their progeny promptly after delivery. Effective treatment requires prompt diagnosis of the illness.

Imaging in the pediatric population presents significant problems due to its small size and unique movements throughout the image acquisition process. This also leads to heightened motion artifacts throughout the acquisition process. This situation poses a technical challenge requiring an elevated spatial resolution, especially in the acquisition of the MRI signal \citep{Ref21}. This situation leads to hesitance among physicians and patients to embrace ML as an alternative to current regimens.

\subsubsection{Data Privacy} 

When patient data is gathered for research, appropriate documentation is executed. The applicants execute consent forms, affirming their agreement to the utilization of data for research or commercial objectives. This technique is also laborious, and most individuals are reluctant to engage in such tasks.

The domain of AI is advancing, although there are increasing ethical apprehensions. Key factors encompass acquiring authorization for data access, safeguarding data security and privacy, mitigating biases and ensuring fairness in algorithms, and fostering transparency \citep{Ref22}. Current laws and regulations do not adequately address the ethical issues raised by the use of AI in the healthcare industry. The formulation of appropriate regulations will require time. We contend that AI is proliferating swiftly, and with this expansion, we can investigate regulations that will guarantee algorithmic openness and data protection.

\subsubsection{Annotated Data Scarcity}

To evaluate a set of CHD algorithms, precise and reliable reference data is essential. Several methods can acquire ground truth data, but the arduous collection and annotation process leads to mistakes and noise in most ground truth annotations. Errors may stem from participants' improper conduct throughout the acquisition process. The acquisition sensor may also influence the data quality.

In complex acquisition scenarios, a practical approach to train and evaluate an ML-based CHD framework is to employ synthetic datasets. The probability of errors is diminished while utilizing synthetic datasets in contrast to those derived from more realistic contexts. Regrettably, the morphology and geometry of the heart are exceedingly intricate. Medical professionals are still unable to fully comprehend the development of the heart and its anomalies. Consequently, the literature is deficient in synthetic datasets for CHD recognition applicable to ML methods.

The annotation of data by healthcare providers presents an additional challenge. Humans predominantly perform the annotation and tagging tasks. One of the most outdated methods for generating ground truth data involves a practitioner assigning a label based on their personal interpretation of CHD. Labelling small datasets is straightforward with this approach; however, it becomes unsuitable for large databases due to the elevated risk of human error. The differing opinions of physicians can result in confusion regarding labelling and annotation. The generation of ground-truth data presents a significant challenge that warrants investigation. Consequently, the literature reports only a limited number of databases.

In the foundational study on CHD recognition, inaccuracies were evident in the ground truth data. The manual procedure is a potential source of these problems. We assert that the generation of ground truth data may not be a leading research domain, yet it is equally vital as any proposed solution for CV applications. Accurate verification and evaluation of any algorithm is unattainable without the prior establishment of ground truth data. Enhanced analysis can solely be attained with more precise ground truth data. The correctness and preparation of the data are contingent upon the precise activity at hand. In the realm of 3D image reconstruction, it is essential to accurately delineate the attributes of the ground truth data for each task. Producing precise data for specific objectives, such gender, ethnicity, and expression classification, is rather uncomplicated. Humans or machines can annotate or automate the labelling process. Producing precise reference data for CHD may prove difficult. Historically, a human conducted manual annotation to produce ground truth. In this process, an individual designates an exact label. Producing precise data using this method is straightforward for smaller databases; nevertheless, as database size increases, it transforms into a tedious and protracted endeavour. Moreover, these methods are more susceptible to human errors. The classification and identification of 35 CHDs pose significant challenges for medical practitioners and neonatologists. This may also explain why the majority of the ground truth data is erroneous.

\subsubsection{Lack of High-Quality Data} 

The calibre of the training data significantly affects how effective most AI systems are. Moreover, the scanned data demonstrate significant quality discrepancies with their elevated expense. The variability in data quality can impede the establishment of a universal network and provide a significant barrier to the commercialization of AI-driven solutions. The datasets accessible to ML specialists for CHD recognition are restricted. We noticed that only 51\% datasets are publically avalable for research. Training AI-based models necessitates an adequate amount of training and testing data. To evaluate and examine inherent biases and overfitting, adequate data is required. Unfortunately, data is not available for download. The diversity in cardiac architecture and the range of distinct disease entities complicate the integration of ML-based solutions in this domain. To mitigate this constraint, comprehensive data from all sources is required, which is unattainable in the foreseeable future. Researchers have yet to explore generative AI and transfer learning as an alternative way to tackle this issue.

\subsection{Results Discussion}

\subsubsection{Features Extraction Methods} 

Current feature extraction techniques present significant challenges for ML specialists in addressing CHD using traditional methods. Preprocessing and feature extraction are critical components in building effective ML systems, and the choice of preprocessing technique depends on the specific characteristics of the dataset. The suitability of a particular technique often varies depending on the type of data being analyzed, as observed across various studies. These methods exhibit considerable variability, and the lack of standardized reporting frameworks adds to the complexity. This variability is partly due to the heterogeneity of CHD-related data; for instance, approaches effective for image data may not be appropriate for audio or tabular data.

\subsubsection{Comparison of Various Classifiers' Performance} 

The diagnosis of CHD has been a long-standing focus of research. Despite limited datasets for training and testing, researchers have achieved notable results. Several classification techniques have been explored, with studies indicating that backpropagation NNs, SVM, LR, and Adaboost often outperform other methods.

In addition, algorithms such as Naive Bayes, RF, KNN, and multilayer perceptron have also been employed with varying degrees of success. The introduction of optimized DL networks has significantly advanced performance, particularly when applied to larger datasets. Enhanced utilization of DL neural networks holds the potential to further optimize classification results, especially with the availability of extensive and diverse data.

\subsubsection{Systems Limitations} 

The quality of training data plays a pivotal role in determining the efficacy of systems used for CHD recognition. Both the training data and the extracted features significantly impact system performance, with superior training data leading to better outcomes. However, most current systems must meet specific criteria to achieve accurate functionality. Failure to adhere to these criteria can result in errors, such as inaccurate disease identification. For instance, overfitting remains a common issue, particularly in DL and traditional ML methods. To address these challenges, researchers should focus on developing adaptive systems that offer greater flexibility in their requirements.

Additionally, generalized approaches must be tailored to suit diverse application scenarios. Improving system efficiency requires a deep understanding of the methodologies and careful utilization of available tools. This approach will help mitigate limitations and enhance the reliability of CHD recognition systems.

\subsubsection{Previous Results Comparison} 

The use of ML for CHD recognition is an evolving area of research that remains underdeveloped. The number of published articles on this topic is relatively limited, and researchers have primarily employed a narrow range of methodologies. Table \ref{yearwise} summarizes the performance outcomes of various approaches over the past eight years, from 2018 to 2024, showing a progressive improvement in results.

An analysis of Table \ref{yearwise} highlights the disparity in performance between TML techniques and contemporary DL methodologies. Data from Table \ref{yearwise} shows that DL techniques consistently outperform TML methods, especially when applied to complex datasets. Interestingly, in some cases, influence-based systems exhibit better performance compared to DL methods. This underscores the importance of developing a more comprehensive understanding of DL algorithms and their applications. Studies such as \citep{Ref13, Ref15} show that DL approaches deliver significantly improved results for complex and extensive databases, further reinforcing their potential. However, Table \ref{yearwise} also reveals varied outcomes for CHD recognition when comparing the effectiveness of traditional ML approaches.

This survey presents a detailed (Table \ref{yearwise}) summarizing data from prior studies utilizing multiple performance criteria, including accuracy, sensitivity, specificity, and others. Nonetheless, we refrained from directly comparing the outcomes of the experimental section due to two reasons. First, the majority of studies utilised disparate datasets for their experiments, complicating the establishment of a meaningful comparison. There is very less repetition of the same database by other authors. Secondly, the absence of comprehensive information concerning dataset configurations, including the particular participants utilized for training and testing, exacerbates the challenge of direct evaluation.

\subsubsection{Data augmentation and synthetic Data}
We anticipate a shift in the recognition of CHD towards innovative DL algorithms. DL methods have training challenges due to insufficient ground truth data. Efficient information transmission could provide a feasible resolution to this issue \citep{Ref180}. It is prudent to explore alternatives such as self-directed learning and supervised learning \citep{Ref181}. Another potential area for improvement involves the utilisation of data augmentation \citep{Ref182} and the implementation of foveated architectural strategies \citep{Ref183}. In DL architectures, data augmentation mitigates the issue of inadequate data.

The adoption of heterogeneous domains is an under-explored area for knowledge transfer. Knowledge transfer is exceptionally proficient in transferring knowledge from the training to the testing phase, particularly when attributes display a certain degree of variability. This substantially decreases the workload necessary for annotating the training data. New developments in DL methods show that we need to learn more about concepts like temporal pooling, and 3D convolution when we process CHD data. While scholars are now examining certain aforementioned methodologies, additional research is required to improve their efficacy for CHD recognition. 

\subsubsection{Vision Transfer Model - To Be Explored for CHD-ML}

DL methodologies have attained exceptional outcomes in diverse imaging assessments. Nonetheless, the challenges of substantial parameter size and diminished throughput restrict their clinical uses. A ViT model is the next thing to be explored with CHD recognition. ViT models can provide numerous benefits for CHD recognition in contrast to conventional ML and DL techniques. ViTs use a self-attention mechanism to find global dependencies in data, which makes them better at finding complex patterns and subtle anomalies that are common in CHD datasets. CNNs, on the other hand, rely on local receptive fields for feature extraction. Moreover, ViTs lack the inductive biases characteristic of CNNs, enabling them to adapt flexibly to many modalities, including echo, X-rays, MRIs, etc. This is essential for CHD recognition, because anatomical variances may be substantial. ViTs scale effectively with large datasets when pre-trained on huge collections and subsequently fine-tuned on specialized medical datasets, facilitating strong feature representation despite insufficient label data. These models can also combine different types of data (for example, imaging data with clinical metadata) to make a more complete diagnosis, which makes the performance of CHD recognition models better. These characteristics render ViT a promising instrument for enhancing CHD detection and diagnosis in clinical environments.

\subsubsection{Gaps in Research}

A major gap exists in the use of AI for the diagnosis and treatment of patients with CHDs across their lifespan. The utilisation of AI in CHD has been restricted by the lack of CHD-specific datasets with labels for model training, the complicated modelling demands caused by diverse clinical characteristics and age-related pathophysiological changes, and the fragmented nature of data within center-specific repositories. Furthermore, at baseline, information regarding particular rare forms of CHDs is limited, requiring multidisciplinary collaboration to compile sufficient datasets. Substantial deficiencies in clinical training, knowledge, experience, and familiarity with AI exist.

CHD continues to be an inadequately explored area for both cardiovascular and ML experts \citep{Ref48,Ref43, Ref38}. The region remains unexplored due to various intricate factors. Hospitals generate significant quantities of data, encompassing clinical information and data from electronic health records. The continuous progression of big data is crucial in healthcare management as it enables the analysis of large datasets to improve illness treatment, determine appropriate therapeutic dosages, and generate predictions \citep{Ref75}. Healthcare produces significant data; nevertheless, a considerable portion remains underused due to challenges in storing, organizing, and understanding complex datasets that often involve multidimensional and nonlinear relationships among variables. The application of these datasets, particularly in uncommon diseases such as CHD, in conjunction with AI predictive models, can assist in identifying individuals at risk of having children with CHD \citep{Ref75,Ref76,Ref77,Ref78}.

\subsection{Some Limitations of the Proposed SLR}

Our SLR has some limitations which we would like to present to the readers. 
\begin{itemize}
    \item First, the diverse characteristics of CHDs and AI methodologies rendered the comparison of individual studies challenging. This SLR primarily concentrated on the overarching constraints of AI in CHD, rather than limitations unique to particular diseases or data types.
    
    \item One limitation of this study is the lack of extensive details concerning specific medical modalities, including echo, X-rays etc. This limitation arises from the authors' lack of medical expertise, which constrains the depth of clinical insights into these modalities. Although we have encapsulated their function in CHD diagnosis and included pertinent findings, a comprehensive analysis necessitates proficiency in medical practice. 
    
    \item Lack of co-authorship or input from medical professionals is missing in the paper. The research presents an AI-centric viewpoint on CHD diagnosis and therapy; nevertheless, the involvement of medical professionals may have enhanced the discourse with more profound clinical insights and better alignment with practical medical applications. Engaging with medical specialists in future research would yield a more thorough and clinically pertinent analysis.

    \end{itemize}
\section{Conclusion}
\label{sec-7}

The integration of AI into CHD research and treatment has brought transformative advancements, demonstrating its potential to revolutionize diagnosis, prognosis, and therapeutic interventions. This comprehensive survey has provided an in-depth review of the AI methodologies applied in CHD recognition, emphasizing the significant progress made and their profound impact on clinical practices.

By analyzing SOA techniques, ranging from TML methods with manually engineered features to advanced DL frameworks, we have outlined the evolution of AI approaches in CHD management. This study also highlights critical challenges, such as the need for diverse and complex datasets to enhance the performance and generalizability of DL models. Furthermore, the curated compilation of publicly available resources, algorithms, and methodologies serves as a valuable reference for fostering future research and innovation in this domain.

The success of AI in CHD care relies heavily on interdisciplinary collaboration. Partnerships among AI researchers, clinicians, bioinformaticians, and industry professionals are essential for translating technological advancements into clinically viable tools. Such collaborations can ensure that AI-driven solutions are tailored to real-world challenges, focusing on patient-centered outcomes and seamless integration into healthcare systems.

AI’s potential for personalized CHD therapy is particularly promising. By leveraging patient-specific data, AI can enable precision treatment plans, optimize therapeutic efficacy, and improve overall patient outcomes. Predictive models driven by AI can minimize trial-and-error approaches, allowing clinicians to deliver more effective and targeted interventions.

While the progress is undeniable, the path to fully realizing AI’s potential in CHD care remains a work in progress. Addressing existing challenges, fostering continuous innovation, and strengthening collaborations are imperative for overcoming current limitations. With sustained efforts, AI has the capacity to redefine CHD management, significantly improving the quality of life for patients and offering hope for a healthier future.

\end{document}